\documentclass[showpacs,preprintnumbers,amsmath,amssymb,twocolumn,aps,pre]{revtex4-1}
\usepackage{default}
\usepackage{amssymb}
\usepackage{color}
\usepackage{xcolor}
\usepackage{amsmath}
\usepackage{mathrsfs}
\usepackage{graphicx}
\usepackage{subfig}
\usepackage{epsfig}

\begin{document}
\title{Bifurcation structure of localized states in the Lugiato-Lefever equation with anomalous dispersion}
\author{P. Parra-Rivas$^{1,2,3}$, D. Gomila$^{2}$, L. Gelens$^{1,3}$ and E. Knobloch$^4$}

\affiliation{ $^1$  Laboratory of Dynamics in Biological Systems, KU Leuven Department of Cellular and Molecular Medicine, University of Leuven, B-3000 Leuven, Belgium\\
$^{2}$Applied Physics Research Group, APHY, Vrije Universiteit Brussel, 1050 Brussels, Belgium\\
$^3$ Instituto de F\'{\i}sica Interdisciplinar y Sistemas Complejos, IFISC (CSIC-UIB), Campus Universitat de les Illes
  Balears, E-07122 Palma de Mallorca, Spain\\
  $^4$Department of Physics, University of California, Berkeley CA 94720, USA}

\date{\today}

\pacs{42.65.-k, 05.45.Jn, 05.45.Vx, 05.45.Xt, 85.60.-q}

\begin{abstract}
The origin, stability and bifurcation structure of different types of bright localized structures described by the Lugiato-Lefever equation is studied. This mean field model describes the nonlinear dynamics of light circulating in fiber cavities and microresonators. In the case of anomalous group velocity dispersion and low values of the intracavity phase detuning these bright states are organized in a homoclinic snaking bifurcation structure. We describe how this bifurcation structure is destroyed when the detuning is increased across a critical value, and determine how a new bifurcation structure known as foliated snaking emerges.
\end{abstract}
\maketitle

\section{Introduction}
Localized dissipative structures (LSs) are found in a large variety of natural systems far from thermodynamical equilibrium including those found in chemistry \cite{chemist}, gas discharges \cite{discharges}, fluid mechanics \cite{fluid}, vegetation and plant ecology \cite{vege}, as well as in optics \cite{spatial_CS}, where they are also known as cavity solitons. These structures arise as a result of a balance between nonlinearity and spatial coupling, and between driving and dissipation. In this work we focus on the field of optics, and study LSs arising in the context of wave-guided cavities such as fiber cavities or microresonators where they are known as temporal solitons \cite{leo_nature}. These systems are described by the Lugiato-Lefever (LL) equation, a mean-field model first introduced in the 80s to describe the transverse component of an electric field in a passive ring cavity partially filled with a nonlinear medium \cite{lugiato_spatial_1987}, and subsequently derived for fiber cavities \cite{
Haelterman}, and microresonators \cite{coen_modeling_2013,chembo_spatiotemporal_2013} as well. In the last few years the LL model has attracted a great deal of interest \cite{specialLLissue} in connection with the formation of frequency combs (FC) in high Q microresonators driven by a continuous wave laser \cite{delhaye_optical_2007,kippenberg_microresonator-based_2011}. FCs correspond to the frequency spectrum of dissipative structures, such as temporal solitons or patterns that circulate inside the cavity. They have been used to measure light frequencies and time intervals with exquisite accuracy, leading to numerous key applications \cite{brasch_2016,okawachi_octave-spanning_2011,papp_microresonator_2014,ferdous_spectral_2011,herr_universal_2012,pfeifle_coherent_2014}. 

In temporal systems bright and dark LSs can be found. Taking into account chromatic dispersion up to second order, two regimes arise, the normal case where the typical structures are dark, and anomalous case where they are bright. The origin and bifurcation structure of dark LSs is well known \cite{Lobanov2015,Parra-Rivas_dark1,Parra-Rivas_dark_long}. However, despite of the large number of publications that yearly address bright states, their origin and bifurcation structure in some regimes is not yet completely understood. The aim of this paper is therefore to present a detailed study of these states. 

Our study is carried out within the LL equation in one extended dimension, namely
\begin{equation}\label{LLE}
 A_t=-(1+i\theta)A+i\nu A_{xx}+i|A|^2A+\rho,
\end{equation}
where $\rho>0$ and $\theta$ are real control parameters representing normalized energy injection and intracavity phase detuning, with $\nu$ taking the value $+1$ for anomalous group velocity dispersion (GVD), and $-1$ for normal GVD. Hereafter we take $\nu=1$ and focus on the anomalous GVD regime.

When the detuning $\theta$ is varied, various features of LSs such as the shape of their tails, width, region of existence, and their dynamics are modified. For low values of the detuning (in particular, for $\theta<2$), we have a good understanding of how different LSs consisting of one or more peaks are organized within the parameter region \cite{Parra-Rivas_PRA_KFCs}. However, current work on frequency combs mostly focuses on higher values of the detuning ($\theta>2$), where the dissipative structures have a sharper profile as well as richer dynamics, including periodic oscillations, temporal chaos, and even spatiotemporal chaos \cite{Leo_OE_2013,Parra-Rivas_PRA_KFCs,godey_stability_2014,Kippenberg_oscillations,Gaeta_oscillations,Anderson_chaos}.

The paper is organized as follows. In Section~\ref{sec:1}, we introduce the time-independent problem, show that it can be recast as a dynamical system in space, and use this system to study time-independent states, such as LSs and periodic patterns. In Section~\ref{sec:2}, we study the linearization of this system to identify the bifurcations from which LSs can emerge. Section~\ref{sec:3} is devoted to weakly nonlinear solutions in the form of LSs in the neighborhood of some of these bifurcations. The bifurcation structure of the different types of LSs arising in our system is analyzed in detail in Section~\ref{sec:4} in the different regimes of interest. The paper concludes in Section~\ref{sec:5} with a discussion of the results.

\section{The stationary problem}\label{sec:1}

In this work we focus on the study of dissipative states which are solutions of the stationary LL equation in the anomalous GVD regime, namely states satisfying the equation
\begin{equation}\label{LLEsta}
i A_{xx} -(1+i\theta)A+i|A|^2A+\rho=0.
\end{equation}
This equation has a number of different solutions, of which the simplest is the homogeneous steady state $A_0\equiv U_0+iV_0$ (hereafter, HSS)
\begin{equation}\label{HSS_real}
\left[\begin{array}{c}
U_0 \\ V_0\end{array}\right]=\left[\begin{array}{c}
\displaystyle\frac{\rho}{1+(I_0-\theta)^2} \\ \displaystyle\frac{(I_0-\theta)\rho}{1+(I_0-\theta)^2}\end{array}\right].
\end{equation}
Here $I_0\equiv|A_0|^2$ satisfies the classic cubic equation of dispersive optical bistability, namely
\begin{equation}\label{HSS}
 I_0^3-2\theta I_0^2+(1+\theta^2)I_0=\rho^2.
\end{equation}
For $\theta<\sqrt{3}$, Eq.~(\ref{HSS}) is single-valued and the system is monostable while for $\theta>\sqrt{3}$ it is triple-valued and the system is then bistable. The latter case is characterized by the presence of a pair of saddle-node bifurcations SN$_{b}$ and SN$_{t}$ located at 
\begin{equation}
 I_{t,b}\equiv|A_{t,b}|^2=\frac{2\theta}{3}\pm\frac{1}{3}\sqrt{\theta^2-3},
\end{equation}
with $I_b\le I_t$ created via a cusp bifurcation at $\theta=\sqrt{3}$. 
In the following we denote the bottom solution branch (from $I_0=0$ to $I_b$) by $A_0^b$,
the middle branch between $I_b$ and $I_t$ by $A_0^m$, and the top branch by $A_0^t$ ($I_0>I_t$).

In addition to HSS the LL equation also possesses stationary solutions we refer to as patterns, and spatially localized solutions or LSs. To explain the difference between these solutions it is useful to employ the language of spatial dynamics. We write the stationary solutions of Eq.~(\ref{LLE}) as a fourth order dynamical system in the variable $x$:
\begin{equation}\label{SD_compact}
\frac{d{\bf y}}{dx}={\bf f}\left({\bf y};\theta,\rho\right).
\end{equation}
Here ${\bf y}(x)\equiv\left[y_1(x),y_2(x),y_3(x),y_4(x)\right]^T$, where $y_1(x)=U(x)$, $y_2(x)=V(x)$, $y_3(x)=dU/dx$, $y_4(x)=dV/dx$, and
${\bf f}({\bf y})\equiv\left[f_1({\bf y}),f_2({\bf y}),f_3({\bf y}),f_4({\bf y})\right]^T$, where
\begin{equation}\begin{array}{l}\label{SD_full}
f_1=y_3\\
f_2=y_4\\
f_3=y_2+\theta y_1-y_1y_2^2-y_1^3\\
f_4=-y_1+\theta y_2-y_2y_1^2-y_2^3+\rho.
\end{array}\end{equation}

This system is invariant under the involution 
\begin{equation}
 R: (x,y_1,y_2,y_3,y_4)\mapsto(-x,y_1,y_2,-y_3,-y_4),
\end{equation}
and is therefore reversible in space. The {\it symmetric section} $\Pi$ is defined as the set of points invariant under $R$, in our case corresponding to the conditions $y_3=y_4=0$, i.e.,
$\Pi\equiv\{(y_1,y_2,y_3,y_4):y_3=y_4=0\}$.
Orbits (including fixed points) closed under the action of $R$ are reversible and intersect $\Pi$.
This symmetry property plays an essential role in the spatial dynamics of the LL equation.

\begin{figure}
\centering
\includegraphics[scale=1]{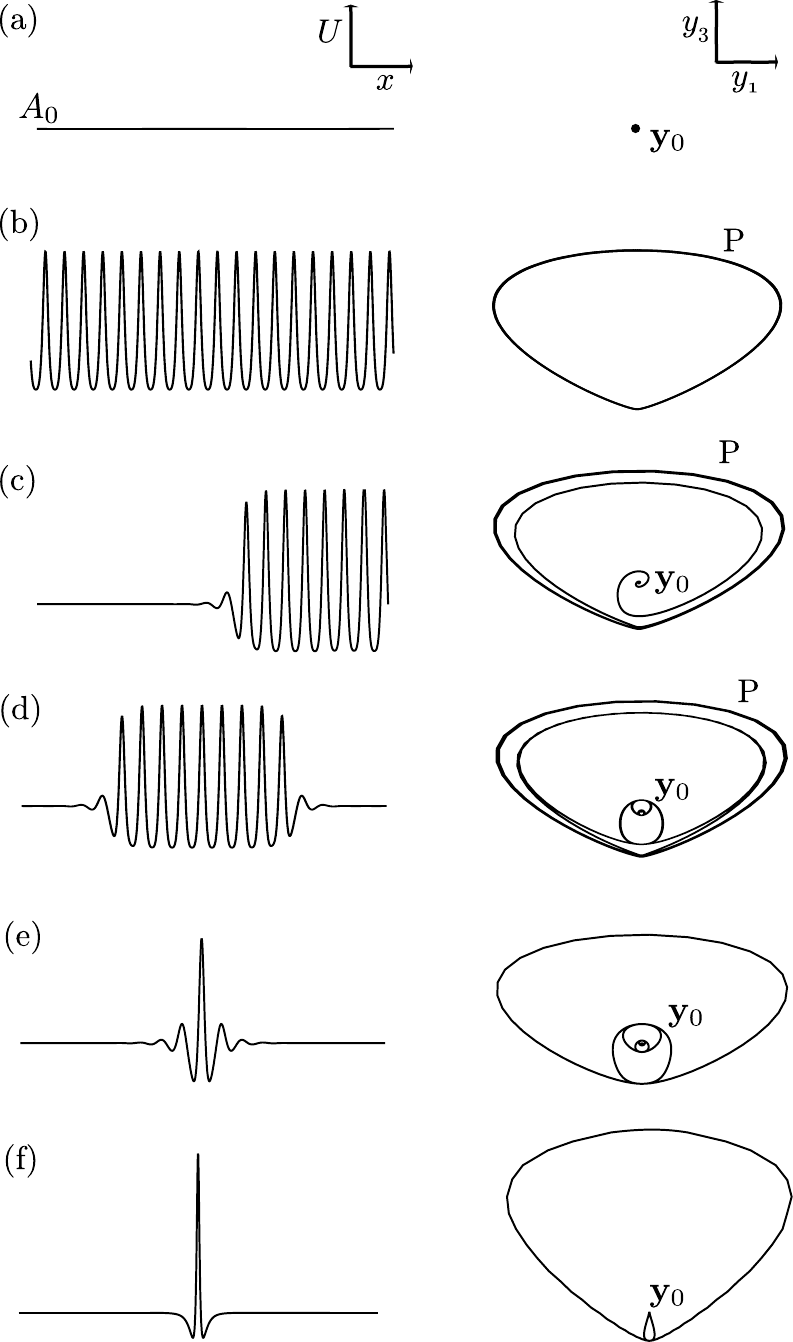}
\captionsetup{justification=raggedright,singlelinecheck=false}
\caption{Possible solutions of Eq.~(\ref{LLEsta}): (a) HSS solution, (b) pattern state, (c) front or domain wall between a HSS and a pattern, (d) LP solution with several peaks, (e) LP solution with a single peak, and (f) single-peak LS in the form of a spike with monotonic tails. The left panels show the field $U(x)$, while the right panels show the corresponding solution of the spatial dynamical system (\ref{SD_compact}) projected on the variables $y_1$ and $y_3$. The correspondence is as follows: (a) fixed point or equilibrium, (b) limit cycle, (c) heteroclinic orbit, (d) and (e) multi-round and single-round wild homoclinic orbits, and (f) tame homoclinic orbit. }
\label{correspondence}
\end{figure}

In the above framework [see Fig.~\ref{correspondence}(a)], the HSS solution $A_0$ corresponds to a fixed point or equilibrium of the system (\ref{SD_full}), namely ${\bf y}_0\equiv(y_{0,1},y_{0,2},y_{0,3},y_{0,4})=(U_0,V_0,0,0)$, while a pattern state corresponds to a limit cycle (periodic orbit) P, as shown in (b). When the state $A_0^b$ coexists with a pattern, a front or domain wall connecting $A_0^b$ with the pattern can form. This interface corresponds to a heteroclinic orbit connecting ${\bf y}_0$ and the periodic orbit P [see panel (c)] that forms as the result of an intersection between the stable manifold of ${\bf y}_0$ and the unstable manifold of P. Spatial reversibility implies that this orbit and its symmetric counterpart intersect forming a heteroclinic cycle \cite{Beck,Knobloch2015}. Homoclinic orbits to ${\bf y}_0$ that bifurcate from this cycle correspond to localized patterns (LPs) containing a long plateau where the solution resembles a spatially periodic pattern as depicted in panel (d) and the corresponding phase space orbit rotates several times about the periodic orbit before returning back to ${\bf y}_0$. Each of these revolutions generates an oscillation (or peak) in the profile of the LP. These types of structures possess oscillatory tails, and the corresponding homoclinic orbits therefore approach and leave ${\bf y}_0$ in an oscillatory manner. A single-peak example of the resulting structure is shown in panel (e). This state is also called an {\it envelope soliton} because of its similarity to a solution of the conservative nonlinear Schr\"odinger equation. These last two orbit types are examples of {\it Shilnikov or wild homoclinic orbits} to a bi-focus fixed point \cite{Sandstede_hom,Campneys_Edgar}, and are both present within the same parameter interval called the pinning or snaking region of the system \cite{Knobloch2015}.
Together with these types of LPs one can find LSs in the form of the single-peak state shown in panel (f) where, in contrast to the previous examples, the tails are monotonic. Such a state corresponds to a homoclinic orbit to a saddle equilibrium, and we refer to it as a {\it tame homoclinic orbit} \cite{Sandstede_hom,Campneys_Edgar} or a {\it spike} \cite{Verschueren}. The formation of this last state is not related to the presence of a pattern, but they arise in global homoclinic bifurcations \cite{Wiggins,Glendining}. Furthermore, such states only appear as isolated peaks and cannot form arrays. These two types of solution bear an intricate relation to one another whose elucidation is one of the primary goals of this paper.

\section{Linearization}\label{sec:2}

The states shown in Fig.~\ref{correspondence} are highly nonlinear and for their study one needs to solve the system (\ref{SD_compact}) numerically. However, the linear regime $A\approx A_0$ allows one to study the way in which the corresponding trajectories leave or approach $A_0$ as $x$ increases and hence provides information about the tail of the solution profiles. This approach allows us to determine the origin of the LP and spike structures, i.e. the bifurcations from which they emerge, and some of their characteristics.

The linearization of Eq.~(\ref{SD_compact}) about the fixed point ${\bf y}_0$ reads
\begin{equation}\label{SD_lin}
 \frac{d}{dx}\delta {\bf y}(x)=\mathcal{D}{\bf f}({\bf y}_0)\cdot\delta {\bf y}(x)+\mathcal{O}(|\delta {\bf y}(x)|^2),
\end{equation}
where $\mathcal{D}{\bf f}({\bf y}_0)$ is the Jacobian
\begin{equation}\label{SPL}
\mathcal{D}{\bf f}({\bf y}_0)=\left[\begin{array}{cccc}
              0&0&1&0\\
               0&0&0&1\\
                \theta-y_2^2-3y_1^2&1-2y_1y_2&0&0\\
                 -(1+2y_1y_2)&\theta-y_1^2-3y_2^2&0&0\\
             \end{array}\right]_{{\bf y}={\bf y}_{0}}.
\end{equation}

\begin{figure}
\centering
\includegraphics[scale=1]{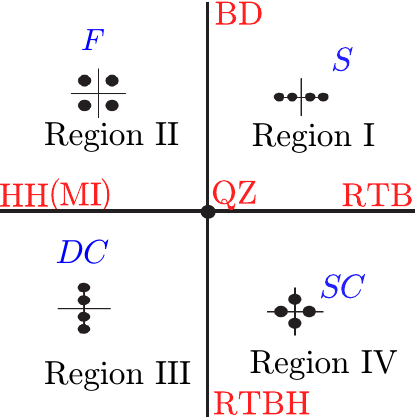}
\caption{(Color online) Sketch of the possible organization of the spatial eigenvalues $\lambda$ satisfying the biquadratic equation (\ref{biqua_general}) for a spatially reversible system. The terminology corresponding to the different regions is explained in Table \ref{tab:template}.}
\label{eigenvalue_regions}
\end{figure}

The solution of the linearized system (\ref{SD_lin}) takes the form $\delta {\bf y}(x)=e^{\mathcal{D}{\bf f}({\bf y}_0)}\cdot\delta {\bf y}_0$, and therefore depends on the eigenvalues of the Jacobian, i.e., the spatial eigenvalues of ${\bf y}_0$. The four eigenvalues of $\mathcal{D}{\bf f}({\bf y}_0)$ satisfy the biquadratic equation
\begin{equation}\label{biqua_general}
 \lambda^4+ c_2\lambda^2+c_0=0,
\end{equation}
with $c_0=\theta^2+3I_0^2-4\theta I_0+1$ and $c_2=4I_0-2\theta$. This equation is invariant under $\lambda\rightarrow-\lambda$ and $\lambda\rightarrow\bar{\lambda}$, and leads to eigenvalue configurations symmetric with respect to both the real and imaginary axes, as depicted in Fig.~\ref{eigenvalue_regions}. The form of this equation is a consequence of spatial reversibility \cite{Devaney_A,Champneys_homoclinic}.

The eigenvalues satisfying Eq.~(\ref{biqua_general}) are
\begin{equation}
 \lambda=\pm\sqrt{(\theta-2I_0)\pm\sqrt{I_0^2-1}}.
\end{equation}
Depending on the control parameters $\theta$ and $\rho$, one can identify four qualitatively different eigenvalue configurations:
\begin{enumerate}
 \item the eigenvalues are real: $\lambda_{1,2}=\pm q_1$, $\lambda_{3,4}=\pm q_2$
 \item there is a quartet of complex eigenvalues: $\lambda_{1,2,3,4}=\pm q_0 \pm ik_0$
 \item the eigenvalues are imaginary: $\lambda_{1,2}=\pm ik_1$, $\lambda_{3,4}=\pm ik_2$
 \item two eigenvalues are real and two imaginary: $\lambda_{1,2}=\pm q_0$, $\lambda_{3,4}=\pm ik_0$
\end{enumerate}

A sketch of these possible eigenvalue configurations is shown in Fig.~\ref{eigenvalue_regions}, and their names and codimension are summarized in Table \ref{tab:template}. The transition from one region to an adjacent one occurs via the following codimension-one bifurcations or transitions:
\begin{itemize}
\item A Belyakov-Devaney (BD) \cite{Devaney_A,Champneys_homoclinic,Haragus} transition occurs between
 regions I and region II. At this point the spatial eigenvalues are real: $\lambda_{1,2}=\pm q_0$, $\lambda_{3,4}=\pm q_0$.
\item The transition between region II and region III is via a Hamiltonian-Hopf (HH) bifurcation \cite{Ioos,Haragus},
with purely imaginary eigenvalues: $\lambda_{1,2}=\pm ik_c$, $\lambda_{3,4}=\pm ik_c$.
\item The transition between region I and region IV is via a reversible Takens-Bogdanov (RTB) bifurcation with 
eigenvalues $\lambda_{1,2}=\pm q_0$, $\lambda_{3}=\lambda_{4}=0$ \cite{Champneys_homoclinic,Haragus}. 
\item The transition between region III and region IV is via a reversible 
Takens-Bogdanov-Hopf (RTBH) bifurcation with eigenvalues $\lambda_{1,2}=\pm ik_0$, $\lambda_{3}=\lambda_{4}=0$ \cite{Champneys_homoclinic,Haragus}.
\end{itemize}
The unfolding of all these scenarios is related to the quadruple zero (QZ) codimension-two point with $\lambda_1=\lambda_2=\lambda_{3}=\lambda_{4}=0$ \cite{Ioos_QZ,Champneys_homoclinic,Haragus,Pere1,Pere2}. In the context of the LL equation these transitions have already been discussed \cite{Parra-Rivas_PRA_KFCs}, but we review the essential findings here as they are needed to understand the changes in the bifurcation structure of LSs that occur as the detuning $\theta$ increases.

In regions I and II of Fig.~\ref{eigenvalue_regions} the equilibrium ${\bf y}_0$ is hyperbolic, i.e. Re$[\lambda]\neq0$, with two-dimensional stable and unstable manifolds. As a result homoclinic orbits to ${\bf y}_0$ are of codimension zero and so persist under small reversible perturbations. In region I, ${\bf y}_0$ is a saddle and the homoclinic orbits are tame. In contrast, in region II, the equilibrium is a bi-focus, and the homoclinic orbits are wild \cite{Sandstede_hom,Campneys_Edgar}. 
\begin{table}[!t]
\centering
\begin{tabular}{|l|c|c|c}
\hline
\hline
Cod & $(\lambda_{1,2,3,4})$  &  Name    \\
\hline
Zero & $(\pm q_0\pm i k_0)$  & Bi-Focus \\
\hline
Zero & $(\pm q_1,\pm q_2)$  & Saddle  \\
\hline
Zero & $(\pm ik_1,\pm ik_2)$  & Double-Center  \\
\hline
Zero & $(\pm q_0,\pm ik_0)$   & Saddle-Center  \\
\hline
One & $(\pm q_0,0,0)$  & Rev. Takens-Bogdanov  \\
\hline
One & $(\pm ik_0,0,0)$ & Rev.Takens-Bogdanov-Hopf    \\
\hline
One & $(\pm q_0,\pm q_0)$  &  Belyakov-Devaney  \\
\hline
One & $(\pm ik_c,\pm ik_c)$  & Hamiltonian-Hopf \\
\hline
Two & $(0,0,0,0)$  & Quadruple Zero  \\
\hline
\hline
\end{tabular}
\captionsetup{justification=raggedright,singlelinecheck=false}
\caption{Nomenclature used to refer to different configurations of the spatial eigenvalues $\lambda$.}
\label{tab:template}
\end{table}
\begin{figure}[!t]
\centering
\includegraphics[scale=1]{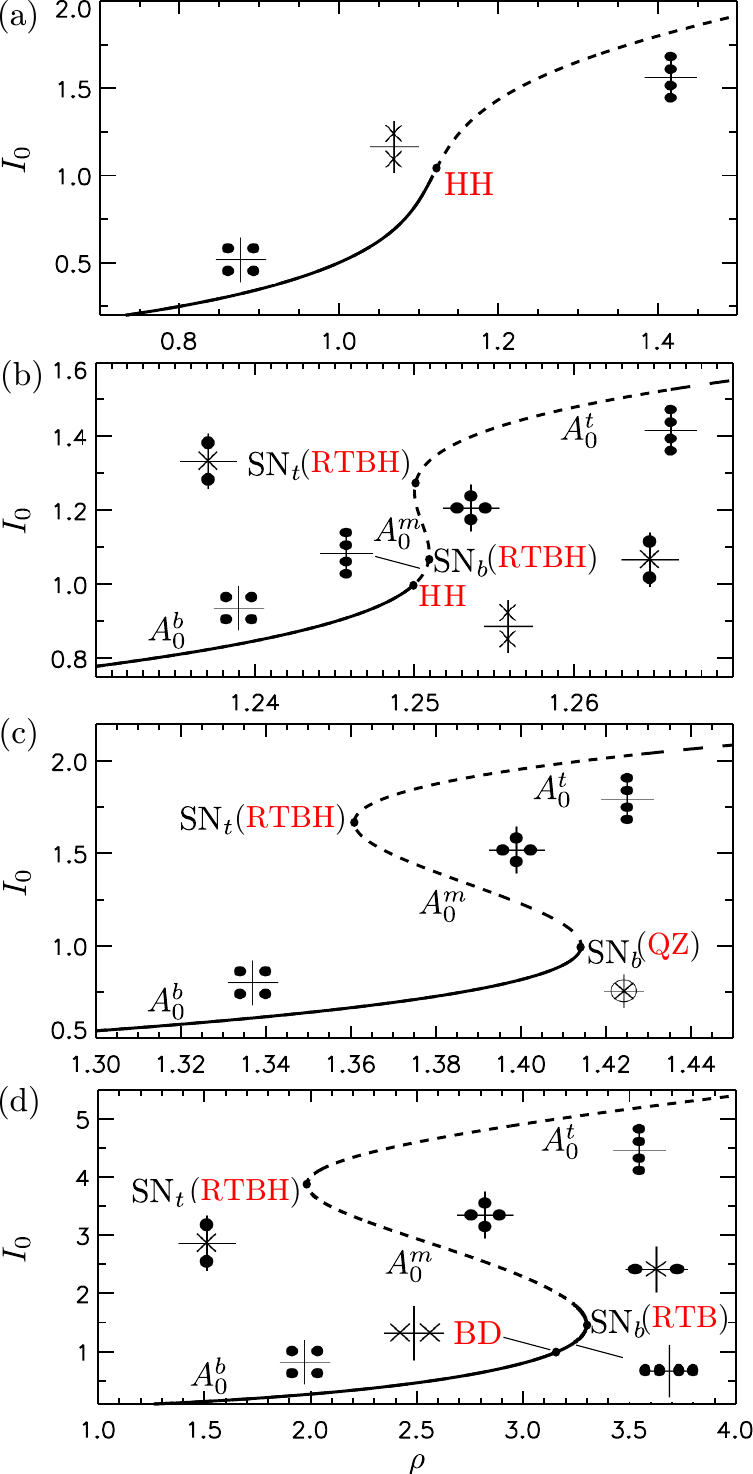}
\caption{(Color online) Homogeneous steady states (HSS) of Eq.~(\ref{LLEsta}) and their spatial eigenvalue configurations for several values of $\theta$. (a) $\theta=1.5<\sqrt{3}$; (b) $\sqrt{3}<\theta=1.75<2$; (c) $\theta=2$; (d) $\theta=4$. Here RTBH stands for the reversible Takens-Bogdanov-Hopf bifurcation, RTB for the reversible Takens-Bogdanov bifurcation and HH for the Hamiltonian-Hopf bifurcation, while BD is the Belyakov-Devaney transition and QZ a quadruple zero codimension-two point. Solid (broken) line indicates temporally stable (unstable) HSS.}
\label{spa_eigen_anomal_full}
\end{figure}
In region IV homoclinic orbits are also possible but only appear at isolated parameter values, or they are homoclinic to small amplitude periodic solutions. This last example corresponds to the so-called {\it generalized solitary waves} \cite{solitary_waves_champneys,Lombardi}.

In our system the condition $I_0=1$ or equivalently
\begin{equation}
 \rho=\rho_c\equiv\sqrt{1+(1-\theta)^2}.
\end{equation}
defines one of the main bifurcation lines in parameter space: for $\theta<2$, this line corresponds to a HH bifurcation in space or equivalently a Turing or modulational instability (MI) in the temporal dynamics that produces small amplitude spatially periodic states. In contrast, for $\theta>2$ it corresponds to a BD transition, a global bifurcation, and no small amplitude states arise.

Two other lines are relevant, corresponding to the saddle-node bifurcations SN$_b$ and SN$_t$, which are defined by $I=I_b$ and $I=I_t$. In the anomalous regime of interest here, $SN_b$ is a RTBH bifurcation if $\theta<2$, and a RTB bifurcation if $\theta>2$. In contrast, SN$_t$ is a RTBH bifurcation for any value of $\theta$.

Figure~\ref{spa_eigen_anomal_full} shows the configuration of the spatial eigenvalues along the HSS solution $A=A_0(\theta,\rho)$ for different values of the detuning $\theta$. In the monostable regime (see panel (a)), $A_0$ is a bi-focus (F) for $I_0<1$  and a double-center (DC) for $I_0>1$. At $\theta=\sqrt{3}$, $A_0$ becomes triple-valued via the cusp bifurcation C, generating three coexisting branches (see panel (b)). The states $A_0^b$ are bi-foci F until $I_0=1$ and DC for $1<I_0<I_b$. Of the remaining HSS the $A_0^t$ are DC, and the $A_0^m$ are saddle-centers (SC) for any value $\theta>\sqrt{3}$. 

As $\theta$ increases HH approaches SN$_b$ resulting in a collision at $(\theta,\rho)=(2,\sqrt{2})$ where $A_0(\theta,\rho)$ undergoes a QZ bifurcation (panel (c)). Finally for $\theta>2$, and as depicted in panel (d), $I_0=1$ is BD transition and the $A_0^b$ are bi-foci F for $I_0<1$, and saddles (S) for $I_0>1$. 

In the present work, we focus on the study of the LSs emerging in the situations depicted in panels (a) and (d). The transition between these scenarios is complex and related to the unfolding of the codimension-two QZ point \cite{Ioos_QZ}. The analysis of the unfolding of this bifurcation point, although essential for the complete understanding of LSs in this system, is beyond the scope of this paper. A study of the LSs arising in region IV of Fig.~\ref{eigenvalue_regions} will be presented elsewhere.

\section{Weakly nonlinear solutions}\label{sec:3}

Normal form theory predicts the existence of small amplitude LSs bifurcating from the HH and the RTB bifurcations \cite{Ioos,Haragus}. In this Section, we use multiscale perturbation theory to compute weakly nonlinear steady states of the LL model in the neighborhood of these bifurcations. These occur at $I_0=I_c$ when $\theta<2$, and at $I_0=I_b$ when $\theta>2$, respectively. The resulting analytical solutions are then used as initial conditions in a numerical continuation algorithm to determine their global bifurcation structure. Following \cite{BuYoKn}, we fix the value of $\theta$ and suppose that the states in the neighborhood of the  bifurcation are captured by the ansatz
\begin{equation}\label{ansatz}
 \left[\begin{array}{c}
U\\V
\end{array}\right]= \left[\begin{array}{c}
U\\V
\end{array}\right]^*+ \left[\begin{array}{c}
u\\v
\end{array}\right],
\end{equation}
where $(U^*,V^*)$ corresponds to the HSS $A^b_0$, and $u$ and $v$ capture the spatial dependence. In each case we introduce appropriate asymptotic expansions for each variable in terms of a small parameter, either $\epsilon=\sqrt{I_0-I_c}$ close to the HH bifurcation or $\epsilon=I_0-I_b$ close to the RTB bifurcation. Here $I_0=|A_0^b|^2$. Note that in the latter case $\epsilon<0$. We use the energy injection $\rho$ as a bifurcation parameter, and therefore write $\rho=\rho_c+\epsilon^2\delta_c$ (close to HH) or $\rho=\rho_b+\epsilon^2\delta_b$ (close to RTB), with
\begin{equation}
 \delta_c=\frac{(\theta-2)^2}{2\rho_c}>0,\label{deltac_HH}
\end{equation}
and 
\begin{equation}
 \delta_b=-\frac{\sqrt{\theta^2-3}}{2\rho_b}<0.\label{deltab_HH}
\end{equation}
Note that $\delta_c$ vanishes at the QZ point $\theta=2$ while $\delta_b$ vanishes at the cusp bifurcation at $\theta=\sqrt{3}$. We now summarize the results of the calculations detailed in the Appendix.


\subsection{Weakly nonlinear states near the Hamiltonian-Hopf bifurcation}\label{sec:3a}

Here $I_0=1$ (i.e., $\rho=\rho_c$) and the appropriate asymptotic expansion for the variables previously defined is
\begin{equation}\label{eq.HSS}
 \left[\begin{array}{c}
U\\V
\end{array}\right]^*= \left[\begin{array}{c}
U_c\\V_c
\end{array}\right]+ \epsilon^2\left[\begin{array}{c}
U_2\\V_2
\end{array}\right]+...
\end{equation}
\begin{equation}
 \left[\begin{array}{c}
u\\v
\end{array}\right]= \epsilon\left[\begin{array}{c}
u_1\\v_1
\end{array}\right]+ \epsilon^2\left[\begin{array}{c}
u_2\\v_2
\end{array}\right]+ \epsilon^3\left[\begin{array}{c}
u_3\\v_3
\end{array}\right]+...,
\end{equation}
where we allow all the variables $u_1,v_1,u_2,v_2,...$ to be functions of $x$ and the long spatial scale $X\equiv\epsilon x$.

\begin{figure*}[!t]
 \centering
  \includegraphics[scale=1]{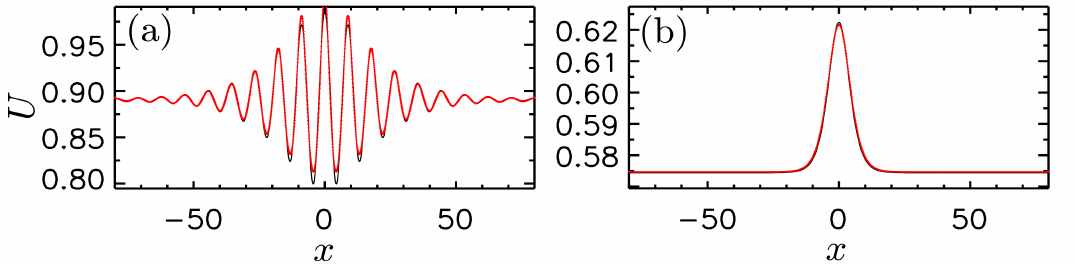}
  \caption{(Color online) Comparison between the asymptotic solutions (black line) in Eqs.~(\ref{new_sech}) and (\ref{bumps_SNh1})
  and the corresponding numerically exact solutions (red line) obtained by numerical continuation. (a) LP with $\theta=1.5$ and $\rho=1.11753$.
  (b) A single spike LS for $\theta=2.5$ and $\rho=1.80460$. In both cases the distance from the instability threshold is $I_{c,b}-I_0=0.036$.}
  \label{comparing_profiles}
\end{figure*}

Inserting these expressions into Eq.~(\ref{LLEsta}) and solving order by order in $\epsilon$,
one finds two types of solution, both given by
\begin{equation}\label{wild_asymp}
\left[\begin{array}{c}
        U\\V
       \end{array}
\right]= \left[\begin{array}{c}
        U_c\\V_c
       \end{array}
\right]+\sqrt{\frac{\rho-\rho_c}{\delta_c}}\left[\begin{array}{c}
        u_1\\v_1
       \end{array}
\right]+\frac{\rho-\rho_c}{\delta_c}\left[\begin{array}{c}
        U_2\\V_2
       \end{array}
\right],
\end{equation}
where $(U_c,V_c)$ is the HSS solution (\ref{HSS_real}) at $I_0=1$ (i.e., $\rho=\rho_c$), $(U_2,V_2)$ represents the leading order correction to this HSS given by
\begin{equation}
 \left[\begin{array}{c}
        U_2\\V_2
       \end{array}
\right]=\frac{\delta_c}{{\left(\theta^{2} - 2 \, \theta +
2\right)} {\left(\theta - 2\right)}}\left[\begin{array}{c}
\theta^{2} \\-\theta^{2} - \theta + 2\end{array}\right],
\end{equation}
and the space-dependent correction is given by
\begin{equation}
 \left[\begin{array}{c}
        u_1\\v_1
       \end{array}
\right]=2B(X) \left[\begin{array}{c}
        a\\1
       \end{array}
\right] {\rm cos}(k_c x+\varphi).
\end{equation}
Here
\begin{equation}
 a=\frac{\theta}{2-\theta}>0
\end{equation}
and $B(X)$ is the solution of the amplitude equation
\begin{equation}\label{amplitude_HH}
C_1 B_{XX}+\delta_c C_2B+C_3B^3=0,
\end{equation}
with coefficients
\begin{equation}
C_1=
-\frac{2 \, {\left(\theta^{2} - 2 \, \theta + 2\right)}}{\theta - 2}>0,
\end{equation}
\begin{equation}       	
C_2=
\frac{2 \, {\left(\theta^{2} - 2 \, \theta +
2\right)}^{\frac{3}{2}}}{{\left(\theta - 2\right)}^{4}}>0,
\end{equation}
and
\begin{equation}      	
C_3=\frac{4 \, {\left(\theta^{2} - 2 \, \theta +
2\right)}^{2} {\left(30 \, \theta - 41\right)}}{9 \, {\left(\theta -
2\right)}^{6}}.
\end{equation}
If the solution of (\ref{amplitude_HH}) does not depend on the long scale $X$, then
\begin{equation}
 B=\sqrt{-\delta_c C_2/ C_3},
\end{equation}
and spatially periodic states (patterns) arise in the form
\begin{multline}\label{new_pattern}
\left[\begin{array}{c}
        U\\V
       \end{array}
\right]= \left[\begin{array}{c}
        U_c\\V_c
       \end{array}
\right]+\left[\begin{array}{c}
        U_2\\V_2
       \end{array}
\right]\frac{\rho-\rho_c}{\delta_c}
\\+2\left[\begin{array}{c}
        a\\1
       \end{array}
\right]\sqrt{\displaystyle\frac{C_2}{C_3}(\rho_c-\rho)}\,\textnormal{cos}\left(k_cx+\varphi\right).
\end{multline}
The pattern can be supercritical or subcritical depending on the value of the detuning $\theta$. 
In particular, $\theta=41/30$ represents the transition from sub- to supercritical where the coefficient $C_3$ vanishes, as predicted in Ref.~\cite{lugiato_spatial_1987}. When $C_3>0$, the pattern bifurcates subcritically, i.e., towards $\rho<\rho_c$; and when $C_3<0$ it bifurcates supercritically, towards $\rho>\rho_c$.

In the subcritical regime, solutions with a large scale modulation, i.e., $X$-dependent solutions, are present. In terms of the original spatial variable $x$ these take the form 
\begin{equation}\label{LS_HH}
 B(x)=\sqrt{\displaystyle\frac{-2\delta_cC_2}{C_3}}\,\textnormal{sech}\left(\displaystyle\sqrt{\frac{C_2(\rho_c-\rho)}{C_1}}x\right),
\end{equation}
resulting in a solution of the form
\begin{multline}\label{new_sech}
\left[\begin{array}{c}
        U\\V
       \end{array}
\right]= \left[\begin{array}{c}
        U_c\\V_c
       \end{array}
\right]+\left[\begin{array}{c}
        U_2\\V_2
       \end{array}
\right]\frac{\rho-\rho_c}{\delta_c}
+2\left[\begin{array}{c}
        a\\1
       \end{array}
\right]\times\\\sqrt{\displaystyle\frac{2C_2}{C_3}(\rho_c-\rho)}\,\textnormal{sech}\left(\displaystyle\sqrt{\frac{C_2(\rho_c-\rho)}{C_1}}x\right)\,\textnormal{cos}\left(k_cx+\varphi\right).
\end{multline}
Evidently these states only exist when $\rho<\rho_c$ and the pattern state is present, i.e., LP states are present when the pattern state bifurcates subcritically but not when it bifurcates supercritically. Within the asymptotic theory the spatial phase $\varphi$ of the background wavetrain remains arbitrary, and there is no locking between the envelope and the underlying wavetrain at any finite order in $\epsilon$. However, this is no longer the case once terms beyond all orders are included \cite{Chapman_Kozy_1,Chapman_Kozy_2,Melbourne}. One finds that generically there are two specific values of the phase $\varphi$ that are selected: $\varphi=0$ and $\varphi=\pi$. These are the only phases that preserve the reversibility symmetry $(x, A)\mapsto(-x, A)$ of Eq.~(\ref{LLE}). The resulting LP states correspond to small amplitude Shilnikov-type homoclinic orbits. A sample solution of this type for $\theta=1.5$, $\rho=1.11753$, and $I_c-I_0=0.036$ is shown in Fig.\ \ref{comparing_profiles}(a), computed both 
from the analytical expression (\ref{new_sech}) (red curve) and from numerical continuation (black curve).

\subsection{Weakly nonlinear states near the reversible Takens-Bogdanov bifurcation}\label{sec:3b}

In this case $I_0=I_b$ (i.e., $\rho=\rho_b$), and the asymptotic expansions read
\begin{equation}\label{eq.HSS_down}
 \left[\begin{array}{c}
U\\V
\end{array}\right]^*= \left[\begin{array}{c}
U_b\\V_b
\end{array}\right]+ \epsilon\left[\begin{array}{c}
U_1\\V_1
\end{array}\right]+ \epsilon^2\left[\begin{array}{c}
U_2\\V_2
\end{array}\right]+...
\end{equation}
for the HSS solution, and 
\begin{equation}
 \left[\begin{array}{c}
u\\v
\end{array}\right]= \epsilon\left[\begin{array}{c}
u_1\\v_1
\end{array}\right]+ \epsilon^2\left[\begin{array}{c}
u_2\\v_2
\end{array}\right]+...
\end{equation}
for the space-dependent terms. This time the fields $u_1,v_1,...$ depend on the long scale $X\equiv\sqrt{\epsilon}x$.

Proceeding in the same fashion as in the previous case, one can establish the existence of weakly nonlinear states around SN$_b$. To first order in $\epsilon$ these are given by
\begin{equation}\label{tame_asymp}
\left[\begin{array}{c}
        U\\V
       \end{array}
\right]= \left[\begin{array}{c}
        U_b\\V_b
       \end{array}
\right]+\sqrt{\frac{\rho-\rho_b}{\delta_b}}\left[\begin{array}{c}
        U_1+u_1\\V_1+v_1
       \end{array}
\right],
\end{equation}
where $(U_b,V_b)$ is the HSS solution (\ref{HSS_real}) at $I_0=I_b$ (i.e., $\rho=\rho_b$) and $(U_1,V_1)$ is the leading order correction to this HSS, namely
\begin{equation}
\left[\begin{array}{c}
U_1 \\ V_1\end{array}\right]=\sqrt{-\delta_b}\mu_b\left[\begin{array}{c}1\\ \eta_b\end{array}\right].
\end{equation}
Here
\begin{eqnarray}
\eta_b &=&-\displaystyle\frac{1}{2}(\theta-I_b-2U_b^2)<0, \\
\mu_b &=& -\sqrt{\frac{-1}{3\eta_b^2V_b+2\eta_b U_b+V_b}}<0.
\end{eqnarray}

The space-dependent contribution is given by
\begin{equation}
\left[\begin{array}{c}
u_1 \\ v_1\end{array}\right]=\left[\begin{array}{c}
U_1 \\ V_1\end{array}\right]\phi(X),
\end{equation}
where $\phi$ is the solution of the amplitude equation:
\begin{equation}\label{psi.eq.1_down}
c_1\phi_{XX}+c_2\phi+c_3\phi^2=0,
\end{equation}
with
\begin{equation}
\begin{array}{ccc}
c_1=-\displaystyle\frac{\mu_b\eta_b}{\sqrt{-\delta_b}}, & c_2=2, & c_3=1.
\end{array}
\end{equation}

Since $\delta_b<0$ and $c_1<0$ when $\theta>2$ (see Appendix) the homoclinic solutions of this equation are 

\begin{equation}
\phi(x)= 
-3\,{\textnormal{sech}}^2
\left[ \displaystyle\frac{1}{2}\sqrt{\frac{2}{\mu_b\eta_b}}(\rho_{b}-\rho)^{1/4}x\right],\label{bumps_SNh1}
\end{equation}
and these are present in $\rho<\rho_b$. These solutions do not exist when $c_1>0$ (i.e., $\theta<2$) when the fold at $A=A_b$ is preceded by HH and $A=A_b$ does not correspond to a RTB bifurcation. Thus Eq.~(\ref{bumps_SNh1}) represents a small amplitude bump on top of the background HSS solution $A_0^b$, i.e., a small amplitude tame homoclinic orbit in the spatial dynamics context. We show an example of this solution in Fig.\ \ref{comparing_profiles}(b) for $\theta=2.5$, $\rho=1.80460$, and $I_b-I_0=0.036$, computed both from the analytical expression (\ref{bumps_SNh1}) (red curve) and from numerical continuation (black curve). The curves are essentially indistinguishable.


\begin{figure}[!t]
\centering
\includegraphics[scale=1.1]{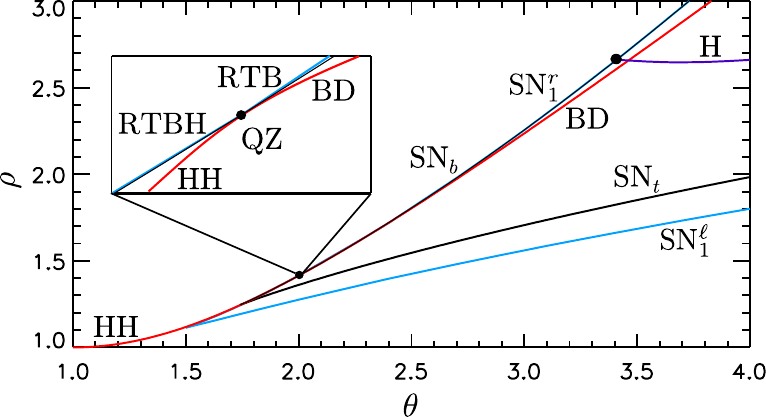}
\caption{(Color online) The parameter plane $(\theta,\rho)$. The red line corresponds to a HH bifurcation for $\theta<2$ but a BD transition for $\theta>2$. The black lines correspond to the saddle-nodes of the HSS $A_0$: SN$_b$ and SN$_t$. In terms of spatial dynamics these lines correspond to the following bifurcations: SN$_t$ is always a RTBH bifurcation; for $\theta<2$ SN$_b$ is also a RTBH bifurcation, but becomes a RTB bifurcation for $\theta>2$. The saddle-nodes of a single peak LS are shown in blue and labeled SN$_1^{\ell}$ and SN$_1^r$; the region inbetween is therefore the region of existence of LSs. The purple line represents a (temporal) Hopf bifurcation (H). The inset shows a close-up view around the QZ point.}
\label{phase}
\end{figure}

\begin{figure*}[t!]
\centering
\includegraphics[scale=1]{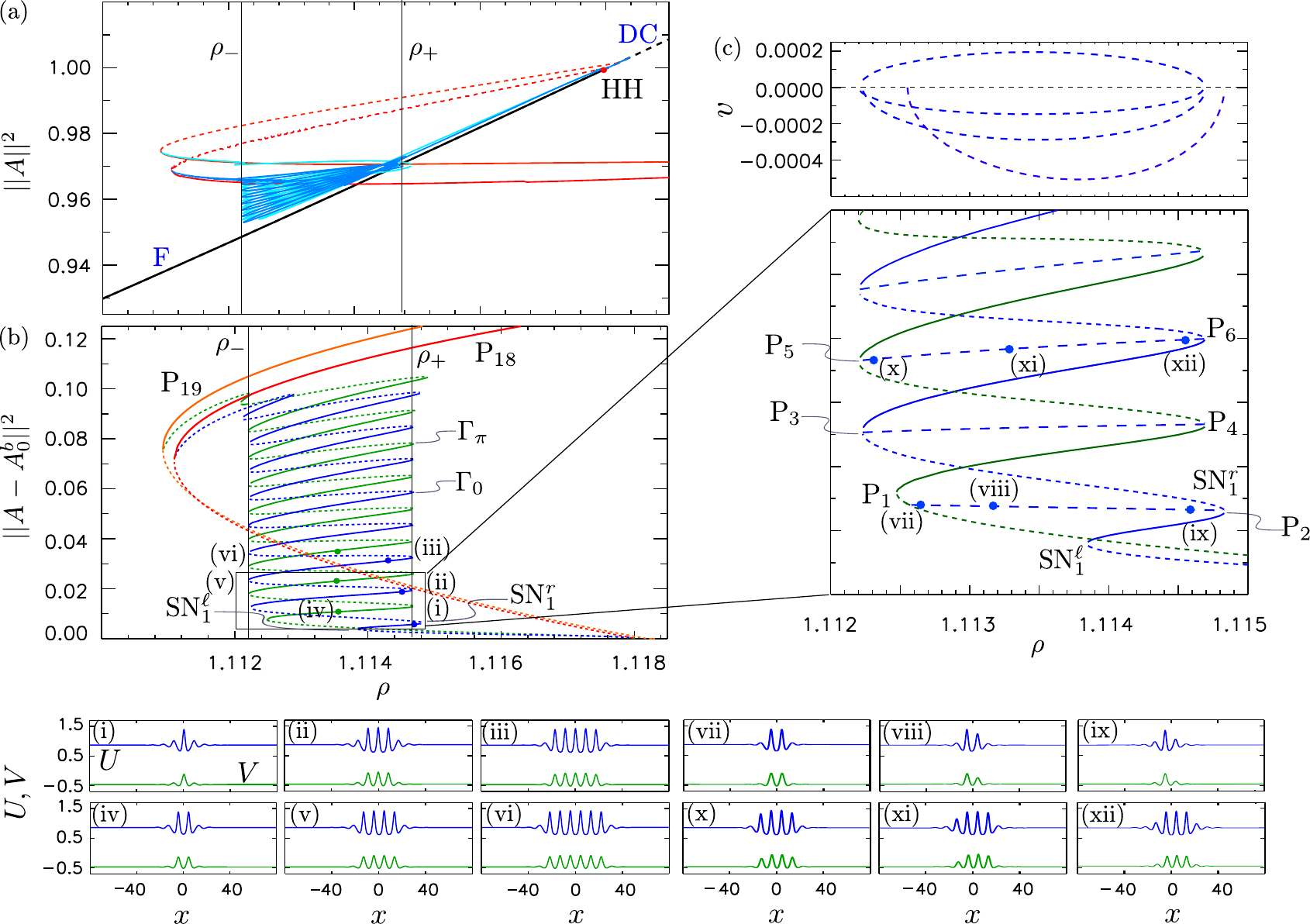}
\caption{(Color online) Homoclinic snaking bifurcation diagram for $\theta=1.5$ and $L=160$. In panels (a) we plot the diagram using $||A||^2$ and in panel (b) we have removed the HSSs $A_0$. Solid (dashed) lines indicate stable (unstable) states. (c) [bottom] Close-up view of the bifurcation diagram in (b) showing the {\it snakes-and-ladders} structure: a pair of intertwined branches of $\varphi=0$ and $\varphi=\pi$ LP states (labeled $\Gamma_0$, $\Gamma_{\pi}$, respectively) with crosslinks consisting of branches of asymmetric states created through a series of Pitchfork bifurcations P$_i$. (c) [top] Drift speed of the rung states as a function of $\rho$ corresponding to the crosslinks ${\rm P}_i{\rm P}_{i+1}$ in the snakes-and-ladders diagram. The drift speed vanishes at the pitchfork bifurcations P$_i$, $i=1,...,4$, and is positive or negative depending on the side at which the extra peak is nucleated as indicated in the $U(x)$, $V(x)$ profiles along the bottom of the figure.}
\label{Diag_theta1.5a}
\end{figure*}

\section{Bifurcation Structure of localized dissipative structures}\label{sec:4}

The weakly nonlinear solutions found in the previous Section are only valid for $\epsilon\rightarrow0$, and therefore, in the neighborhood of the bifurcation. In this Section, we apply numerical continuation algorithms to track these solutions to parameter values far from the bifurcation. This procedure allows us to determine the different solution branches and their temporal stability as a function of the control parameters of the system, and hence their region of existence. For $\theta<2$, LSs bifurcating from HH undergo {\it homoclinic snaking}, as shown previously in both the LL equation \cite{Gomila_Schroggi,Parra-Rivas_PRA_KFCs} and in other systems \cite{Woods1999,Coullet, Burke_Knobloch,Knobloch2015}. Here we review the main features of this structure and compute the so-called rung states \cite{Burke_ladders}. For $\theta>2$ we find instead that the bump state that exists close to the SN$_b$ undergoes a different type of bifurcation structure, which is morphologically 
equivalent to the {\it foliated snaking} found in \cite{Ponedel_Knobloch}. This new structure coexists with disconnected remnants of the homoclinic snaking branches. We show that this disconnection, and therefore the collapse of the homoclinic snaking, is the consequence of a global bifurcation that takes place at the BD transition \cite{Champneys_homoclinic}.

Figure~\ref{phase} shows the parameter plane $(\theta,\rho)$ with the main bifurcation lines labeled. The region between the blue lines SN$_1^{\ell}$ and SN$_1^r$ corresponds to the region of existence of single peak LSs. The purple line H corresponds to a (temporal) Hopf bifurcation that arises in a Fold-Hopf bifurcation \cite{Holmes}. Above this line the LSs start to oscillate and can exhibit both temporal and spatiotemporal chaos \cite{Leo_OE_2013,Parra-Rivas_PRA_KFCs,godey_stability_2014,Kippenberg_oscillations,Gaeta_oscillations,Anderson_chaos}. In this work we focus on the region where the LSs are stationary.

\subsection{Homoclinic snaking structure}\label{sec:4a}

When $41/30<\theta<2$, the HH bifurcation gives rise to a pair of weakly nonlinear LP states corresponding to $\varphi=0$ and $\varphi=\pi$ [see Eq.~(\ref{LS_HH})], simultaneously with a spatially extended pattern with wavenumber $k_c$. Figure~\ref{Diag_theta1.5a} shows the bifurcation structure of these states on a domain size $L=160$ as the parameter $\rho$ is varied for a fixed value of detuning $\theta=1.5$.

In panel (a), we plot the energy (i.e., the $L^2$ norm)
\begin{equation}
 ||A||^2=\displaystyle\frac{1}{L}\int_{-L/2}^{L/2}|A(x)|^2\,dx.
\end{equation}
as a function of the bifurcation parameter $\rho$. 
Panel (b) shows the same diagram but with the background field $A_0^b$ removed, a representation that opens up the snaking behavior
that is hard to discern in panel (a).

Two branches of LP solutions are found: one branch is associated with the phase $\varphi=0$ (in blue) and corresponds to profiles with local maxima in $A(x)$ at the midpoint $x=0$ [see subpanels (i)-(iii)]; the other branch (in green) is associated with the phase $\varphi=\pi$ and corresponds to profiles with local minima in $A(x)$ at $x=0$ [panels (iv)-(vi)]. We refer to the former branch as $\Gamma_0$ and the latter as $\Gamma_{\pi}$. All these structures consist in a slug of the pattern state embedded in a homogeneous state. Both branches emerge subcritically from HH and persist to finite amplitude where they undergo homoclinic snaking, i.e., a sequence of back-and-forth oscillations that reflect the successive addition of a pair of wavelengths, one on each side of the structure, as one follows $\Gamma_0$ and $\Gamma_{\pi}$ upwards. These take place within the parameter interval $\rho_-<\rho<\rho_+$ determined by the first and last tangencies between the unstable and stable manifolds of $A_0^b$ and the periodic state, and called the snaking or pinning region \cite{Woods1999,Burke_Knobloch,Knobloch2015}. The folds or saddle-node bifurcations on either side converge monotonically and exponentially rapidly to $\rho_-$ and $\rho_+$, both from the right. Within the snaking or pinning region one therefore finds an infinite multiplicity of LPs of different lengths. In the present case many of these are (temporally) stable, as indicated by the solid lines in the figure. In finite domains the wavelength adding process must terminate, of course, and in periodic domains one finds that both snaking branches turn over and terminate near the fold on one of the many branches of periodic states that are present \cite{Bergeon}. In the present instance the LP branches in fact terminate on different branches of periodic states. This is a finite size effect and is well understood.
\begin{figure*}[t!]
\centering
\includegraphics[scale=1]{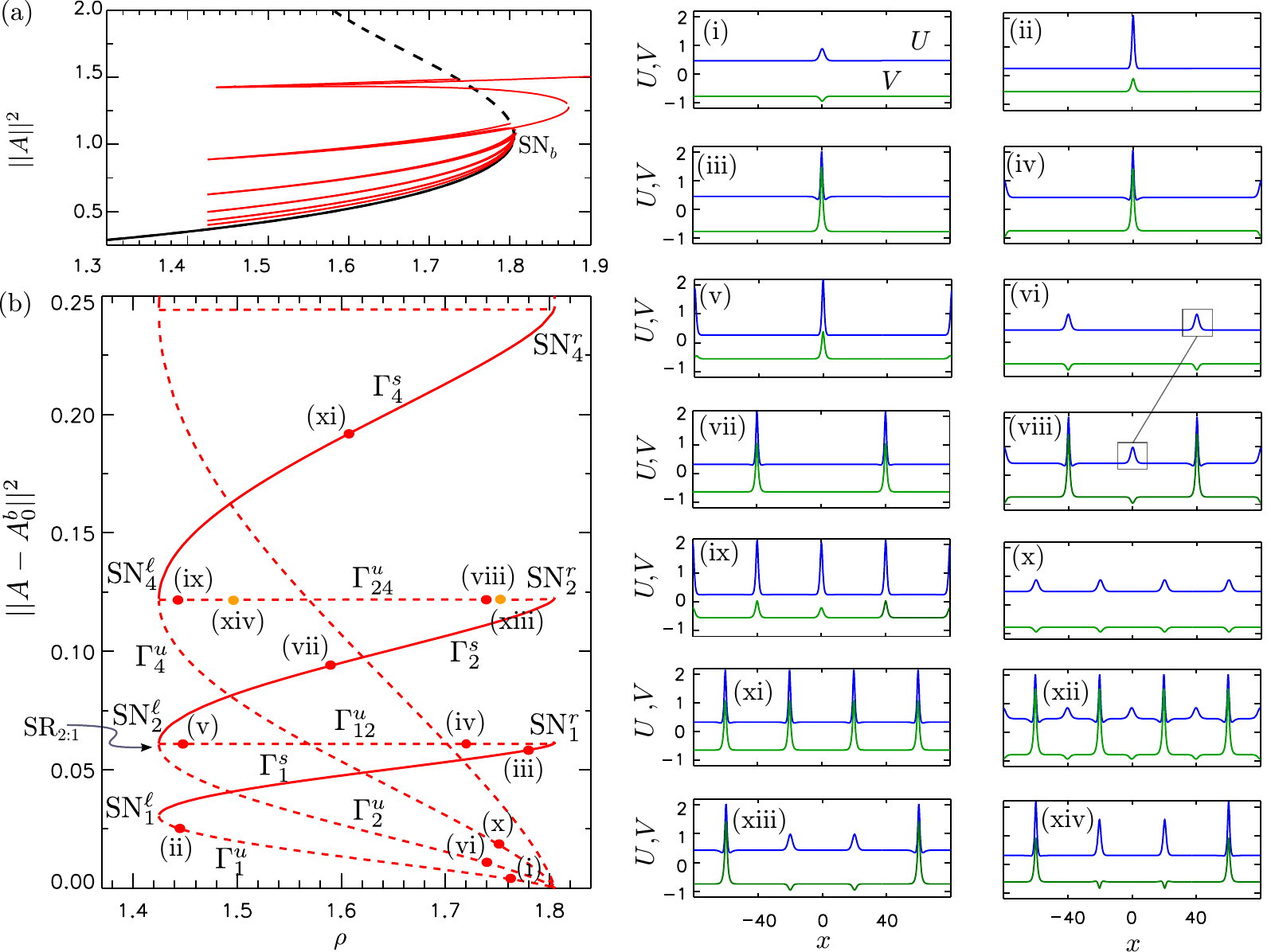}
\caption{(Color online) (a) Bifurcation diagram for $\theta=2.5$ showing the norm $||A||^2$ as a function of $\rho$. (b) The same diagram but in terms of $||A-A_0^b||^2$ revealing the foliated snaking nature of the diagram; the numbered dots correspond to the profiles shown in the panels on the right. On the real real line the different branches of equispaced multi-peak states all bifurcate from the RTB point at SN$_b$. All branches turn around in the vicinity of $\rho_b\approx 1.80501$ where the small peaks first appear as $\rho$ decreases. As shown by the boxed peaks in profiles (vi) and (viii) the secondary peaks that appear along $\Gamma_{12}^u$ are identical to the small peaks along the lower branch $\Gamma_1^u$ and similarly for the other branches. These secondary peaks grow with decreasing $\rho$ allowing the branch to terminate on a branch with twice as many peaks in a 2:1 spatial resonance located near the saddle-node SN$_{2n}$. The yellow dots, corresponding to profiles (xiii) and (xiv), show that 
branches in which the small and large peaks are ordered differently are also present (compare these profiles with (viii) and (ix)). The corresponding branches are indistinguishable in panel (b) since the quantity shown, $||A-A_0^b||^2$, is insensitive to the ordering of the peaks. In (b) solid (dashed) lines represent stable (unstable) branches. Stability is also indicated using the superscripts $s,u$.}
\label{Diag_theta2.5b}
\end{figure*}

The formation of these LPs, and their organization in a homoclinic snaking structure, can be understood in terms of a {\it heteroclinic tangle} that forms within $\rho_-<\rho<\rho_+$ as a result of the transversal intersection of the unstable manifold of $A_0^b$ [$W^u(A_0^b)$] and the stable manifold of a given periodic pattern P [$W^s(P)$] as $\rho$ varies \cite{Woods1999,Coullet,Beck}. The first tangency between $W^u(A_0^b)$ and $W^s(P)$ at $\rho_-$ corresponds to the birth of Shilnikov-type homoclinic orbits biasymptotic to the bi-focus equilibrium $A_0^b$ and the last tangency at $\rho_+$ corresponds to their destruction \cite{Gomila_Schroggi}. The asymmetric rung states that form an important part of the {\it snake-and-ladders} structure of the snaking or pinning region \cite{Burke_ladders} form as a result of secondary intersections between $W^u(A_0^b)$ with $W^s(A_0^b)$ that take place outside of the symmetry plane $\Pi$ \cite{Beck,Makrides}. The branches corresponding to these states are shown in the close-up view in panel (c) [bottom]. These states are all unstable [see panels (vii)-(ix)] and arise in pitchfork bifurcations located near the saddle-node bifurcations on the $\Gamma_0$ and $\Gamma_{\pi}$ branches. Consequently, each rung in the figure corresponds to two states related by reflection symmetry, and hence, of identical $L^2$ norm. Since the LL equation does not have gradient dynamics, any asymmetric state will necessarily drift. The drift speed $v$ depends on the control parameters of the system, as one can see in panel (c) [top] which shows $v$ as a function of $\rho$. Profiles (vii) to (ix), on branch P$_1$P$_2$, show the evolution of asymmetry with decreasing $\rho$ and the gradual transformation from a $\Gamma_0$ LP state to a $\Gamma_{\pi}$ LP state. The velocity of these states is negative and vanishes at the pitchfork bifurcations at either end of the branch. In addition to these states, there are also branches of asymmetric rung states where the extra peak grows on the left of the initial peak. An example of such states is shown in panels (x)-(xii) taken from branch P$_5$P$_6$. The velocity of these states is positive as shown in Fig.~\ref{Diag_theta1.5a}(c). When higher-order dispersion terms are included, breaking the $x$-reversibility of Eq.~(\ref{LLEsta}), the pitchfork bifurcations become imperfect resulting in the break-up of the snakes-and-ladders structure into a stack of isolas \cite{Burke_breaking,Parra_Rivas_3}.

In addition to the above {\it single pulse} states one also finds bound states of LPs, i.e., {\it multipulse} states, that form as a result of the locking of two or more LPs at distances given by half-integer multiples of the wavelength $\Lambda=2\pi/{\rm Im}[\lambda]$ \cite{Burke2009,Lloyd2011,PPR_interaction} owing to the presence of oscillatory tails in all LPs present for $\theta<2$.

\begin{figure}[t!]
 \centering
  \includegraphics[scale=1]{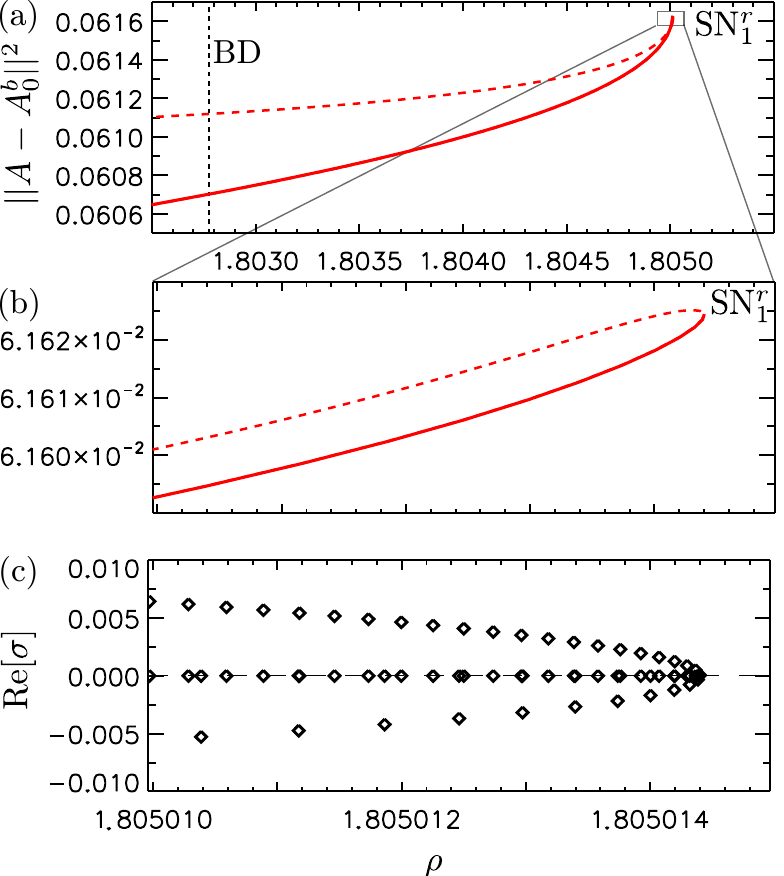}
 \caption{(Color online) (a)-(b) Detail of the saddle-node SN$^r_1$. (c) Real part of the temporal eigenvalue $\sigma$ asociated to this fold. The zero eigenvalue is a consequence of translation invariance.}
  \label{detail}
\end{figure}

\subsection{Foliated snaking}\label{sec:4b}

\begin{figure*}[t!]
\centering
\includegraphics[scale=0.946]{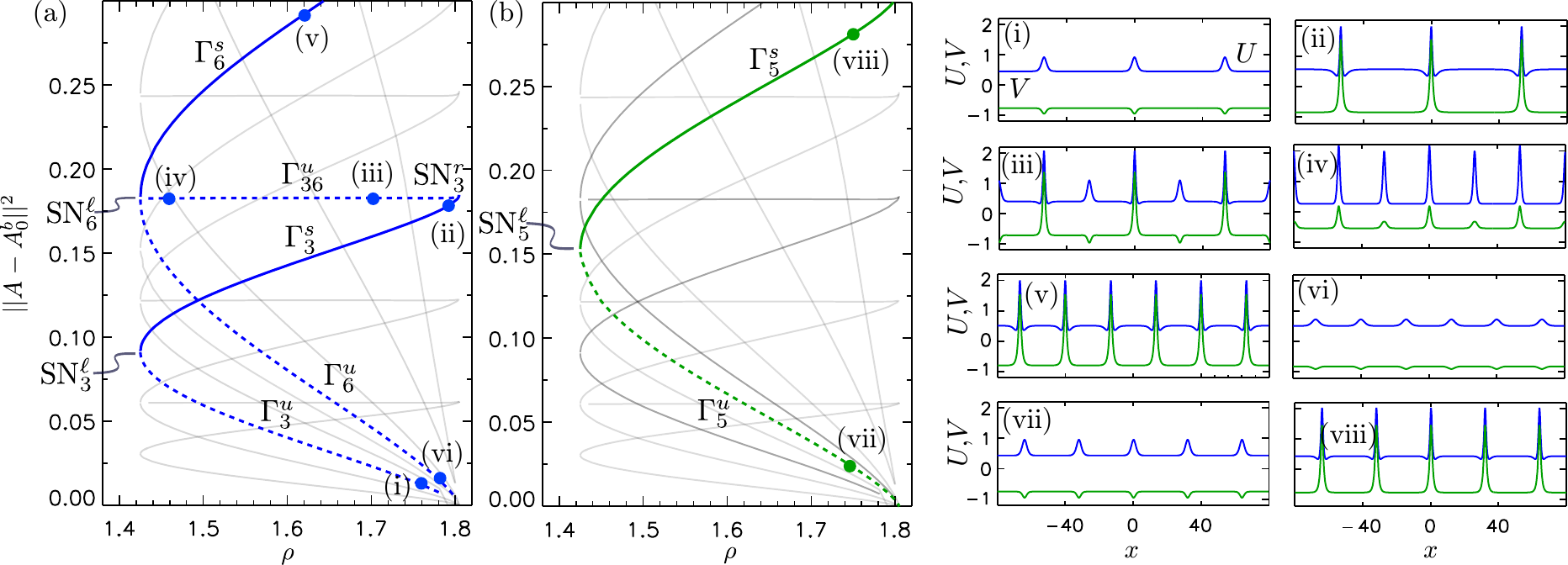}
\caption{(Color online) (a) Branches of equispaced three-peak and six-peak states (blue curves) corresponding to the states shown in panels (i)-(vi); as in Fig.~\ref{Diag_theta2.5b} branches corresponding to states in which the small and large peaks are ordered differently coincide with $\Gamma_{36}^u$. (b) Branch of equispaced five-peak states (green curve) associated with states (vii)-(viii). In both panels the gray branches in the background represent those already shown in Fig.~\ref{Diag_theta2.5b}. In (a) solid (dashed) lines represent stable (unstable) branches.}
\label{dia_25_extra}
\end{figure*}

When $\theta>2$ the HH bifurcation is replaced by a BD transition, and no small amplitude dissipative structures emerge from it. In this situation, $A_0^b$
is stable all the way until SN$_b$. This point corresponds to a RTB in the context of spatial dynamics and small amplitude LSs do emerge from such a point 
(see Section~\ref{sec:3}). This small amplitude LS can then be tracked as $\rho$ decreases and the resulting highly nonlinear states calculated. The type
of bifurcation diagram one obtains for $\theta=2.5$ is shown in Fig.~\ref{Diag_theta2.5b}(a). As before the diagram becomes clearer when the background 
state $A_0^b$ is removed from the norm plotted along the vertical axis, as shown in panel (b). The solution profiles $U(x)$, $V(x)$ corresponding to the 
various branches are shown in the panels on the right of the figure. On the real line the RTB bifurcation corresponding to the fold SN$_b$ ($\rho_b=1.80501$)
is responsible for the appearance of multiple branches of spatially localized states. All of these have monotonic tails and so interact weakly. Profile (i)
shows a single peak state in the available domain soon after it bifurcates from the vicinity of SN$_b$. As one proceeds up the corresponding
branch $\Gamma_{1}^u$, that is $\rho$ decreases, the central peak grows in amplitude [profile (ii)] until the fold SN$_1^{\ell}$ where this
state acquires stability. The peak continues to grow along the subsequent stable branch [profile (iii)] until SN$_1^r$. Figure \ref{detail} shows
the behavior near this point, and demonstrates that the cusp-like structures in Fig.~\ref{Diag_theta2.5b}(b) at $\rho\approx\rho_b=1.80501$ are
in fact regular folds, with a zero temporal eigenvalue at the fold (i.e., a change in the stability) as expected. Beyond this point, along branch 
$\Gamma_{12}^u$, the solution develops a subsidiary peak located at $x=L/2$ [equivalently $x=-L/2$, profile (iv)] and this peak grows 
to the amplitude of the original peak by the time the branch reaches SN$_2^{\ell}$ on the left and terminates on the branch $\Gamma_{2}^u$ of
equispaced two-peak states within the periodic domain $[0,L)$ [profile (v)]. The termination point thus corresponds to a 2:1 spatial 
resonance \cite{Armbruster,Porter,Proctor} that occurs at the point SR$_{2:1}$ close to this fold.
The two-peak state likewise originates near SN$_b$ [profile (vi)] and undergoes similar behavior to that of the single peak state. 
Specifically, it acquires stability above SN$_2^{\ell}$ [profile (vii)] and terminates at SN$_2^r$. Beyond this point,
along branch $\Gamma_{24}$, intermediate peaks appear midway between the large peaks already present [profile (viii)], and these grow to full 
amplitude by the time the next SR$_{2:1}$ point is reached near SN$_4^{\ell}$, and the branch terminates on the branch of equispaced four-peak
states [profile (ix)]. This state likewise originates in the vicinity of SN$_b$ [profiles (x) and (xi)]. This process repeats, resulting in a 
cascade of equispaced states with $2^n$ peaks.

Figure \ref{Diag_theta2.5b} also shows two additional profiles, labeled (xiii) and (xiv), in which the small and large peaks are ordered differently (compare these profiles with (viii) and (ix)). The corresponding branches are indistinguishable in panel (b) since the quantity shown, $||A-A_0^b||^2$, is insensitive to the ordering of the peaks. Indeed, we expect branches corresponding to all possible equispaced orderings of small and large peaks to be present. In addition, we expect (but do not show) that branches with different numbers $n_{\rm small}$, $n_{\rm large}$ of small and large peaks ($n_{\rm small}+n_{\rm large}=n$) also bifurcate from the vicinity of each SN$_n^{\ell}$ as discussed in greater detail in Ref.~\cite{Lojacono}. Moreover, the boxed peaks in profiles (vi) and (viii) show that the small peaks that appear along branch $\Gamma_{12}^u$ are nothing but the small peaks present at the corresponding $\rho$ value along the lower branch $\Gamma_1^u$, and similarly for all the other branches. 
These properties are identical to those associated with foliated snaking as described in Ref.~\cite{Ponedel_Knobloch} showing that the structure identified in the present problem is in fact likely to be universal in these types of problems.
\begin{figure}[!h]
 \centering
  \includegraphics[scale=1]{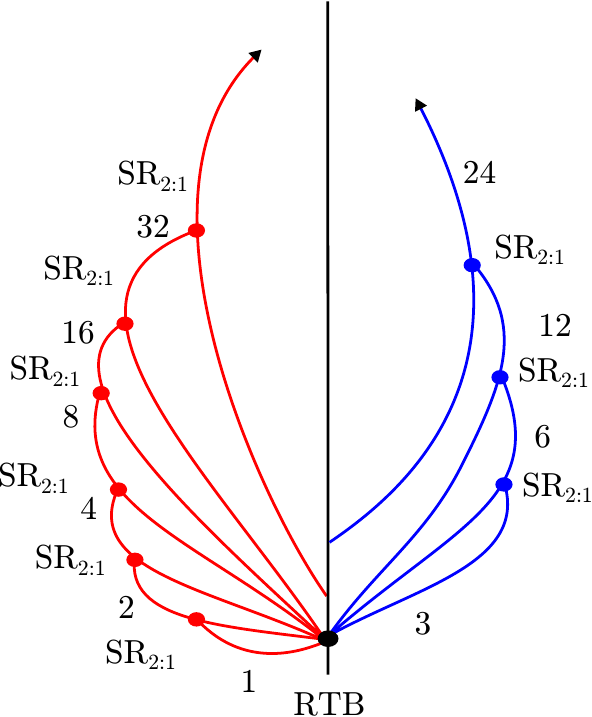}
 \caption{(Color online) Schematic bifurcation diagram showing the topological features of foliated snaking. The red branches correspond to Fig.~\ref{Diag_theta2.5b} while the blue ones correspond to Fig.~\ref{dia_25_extra}. The vertical line represents the HSS. On the real line all branches bifurcate from the RTB bifurcation at SN$_b$ generating a large multiplicity of stationary localized states only some of which are shown in the figure. This scheme is inspired by a similar one appearing in Ref.~\cite{Ponedel_Knobloch}. }
  \label{esquemaf1}
\end{figure}


On top of this basic skeleton, one can find similar bifurcation structures for states with $n$ equispaced peaks, where $n$ is any integer. 
Two examples are shown in Fig.~\ref{dia_25_extra} with $n=3$ [panel (a), blue curves] and $n=5$ [panel (b), green curves]. 
The branches already described are shown in the background using gray lines. The profiles corresponding to panel (a) are shown on 
the right [profiles (i)-(vi)] while those corresponding to panel (b) are shown as profiles (vii)-(viii). We see that the branch 
of three equispaced peaks acquires stability at SN$_3^{\ell}$ on the left and terminates in a cusp-like structure at SN$_3^r$ on the right.
Thereafter subsidiary peaks develop generating the branch $\Gamma_{36}^u$ that terminates in its own 2:1 
spatial resonance near the saddle-node SN$_3^{\ell}$. The branch of equispaced five-peak states exhibits the same behavior,
as do all the other branches for which $n$ is an odd integer or of the form $p2^q$, where $p$ is odd.


The connectivity of the solution branches described above differs from the typical homoclinic snaking studied in Section~\ref{sec:4a}, and resembles so-called {\it foliated snaking} identified in Ref.~\cite{Ponedel_Knobloch} and illustrated schematically in Fig.~\ref{esquemaf1}. In this work, however, this structure was generated by externally imposed spatially periodic forcing. In our case the same structure is generated intrinsically and is a consequence of a global bifurcation that takes place at the BD transition.

\subsection{Collapse of Homoclinic Snaking}\label{sec:4c}

The foliated snaking identified in the previous section only relates states consisting of $n$ equispaced but isolated LS each within the same periodic domain of length $L$. The structures that form are therefore periodic arrays of identical peaks with different wavelengths. In this section we explore the connection of these states with the LS states found when $41/30<\theta<2$ that possess oscillatory tails and undergo standard homoclinic snaking as a result of transverse intersection of stable and unstable manifolds of HSS and the spatially extended periodic state P. For $\theta>2$, heteroclinic tangles between $A_0^b$ and subcritical patterns can still arise, and therefore LPs can still be present. Nevertheless, these type of LSs, and the pattern solution involved in their formation, do not emerge from small amplitude states such as those present near the HH point, but arise instead from a global homoclinic bifurcation \cite{Wiggins,Glendining}.
\begin{figure}[!t]
\centering
\includegraphics[scale=0.96]{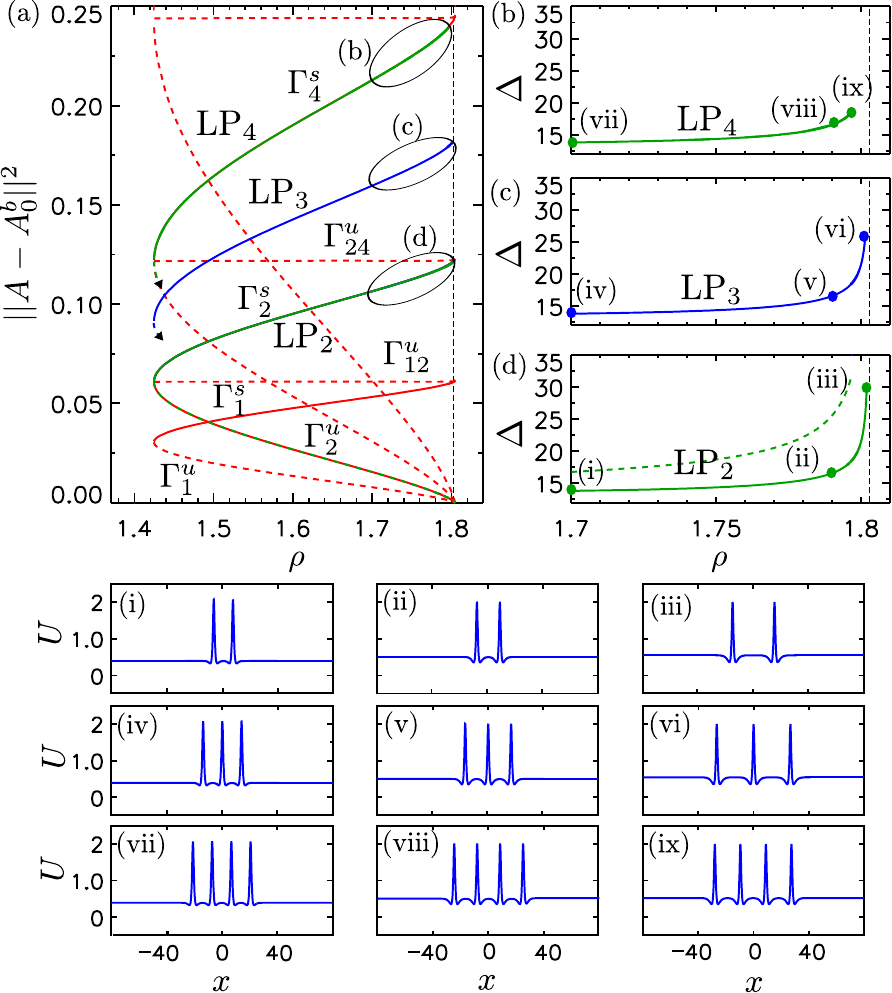}
\caption{(Color online) Branches of LP states for $\theta=2.5$ consisting of clumps of peaks. Panel (a) shows LP branches connected by homotopy with those undergoing homoclinic snaking at $\theta=1.5$ (green and blue curves). The solution branches corresponding to the foliated snaking shown in Fig.~\ref{Diag_theta2.5b} are shown in red. Panels (b)-(d) show, respectively, LP states consisting of 2-peak, 3-peak and 4-peak clumps. The panels show the separation $\Delta$ between peaks within the clumps as a function of $\rho$ along these branches. The separation $\Delta$ diverges as $\rho\rightarrow\rho_{\rm BD}$ (dashed vertical lines) and at this point a global bifurcation takes place. The clumped states fall on top of the evenly spaced states in the bifurcation diagram (red curves in panel (a)) because the norm $||A-A^b_0||^2$ only counts the number of peaks.
}
\label{sepa}
\end{figure}

The aim of this Section is to clarify the origin and type of bifurcation structure that these LPs undergo for $\theta>2$. As we shall see, these states follow the remnants of a homoclinic snaking structure whose solution branches disconnect abruptly at the BD point and reconnect with branches organized within the foliated snaking structure.

The LP states present for $\theta<2$ can be continued by suitably varying both $\theta$ and $\rho$ [see Fig.~\ref{phase}] into the region $\theta>2$. In particular, we have continued numerically LPs with two, three and four peaks, such as those shown in Fig.~\ref{Diag_theta1.5a}, from $\theta=1.5$ to $\theta=2.5$. Once $\theta=2.5$ is reached the parameter $\rho$ can be varied and the corresponding solution branch traced out. These branches are shown in Fig.~\ref{sepa}(a). The branches related by homotopy with the $\Gamma_0$ family of LPs are shown in blue, while those related with the $\Gamma_{\pi}$ branch are shown in green. The profiles of the resulting states at different values of $\rho$ are shown in the subpanels at the bottom and labeled (i)-(ix); these show that for $\theta>2$ the LP states consist of clumps of peaks with interpeak separation given by the wavelength of one of the coexisting periodic states. The latter can, in principle, be obtained by continuation from $\theta<2$ but there is in 
general a one-parameter family of such states within a wavenumber interval called the Busse balloon whose amplitude depends only weakly on $\rho$. In contrast, the LP wavelength is selected uniquely by the fronts bounding the structure on either side and depends strongly on $\rho$, i.e. the presence of the fronts leads to a particular cut through $(\rho,k)$ space \cite{BurkeMaKnobloch}, and it is along this cut that $k\to 0$ or equivalently the LP wavelength diverges. For LP states consisting of many peaks this limit must correspond to the BD point. These conclusions are consistent with our numerical tracking of 2-, 3-, and 4-peak LPs from $\theta<2$, where they are formed via homoclinic snaking, into $\theta>2$ and then through the BD point into the foliated snaking region. For comparison, Fig.~\ref{sepa}(a) also shows the branches taking part in the foliated snaking from Fig.~\ref{Diag_theta2.5b} (red curves); these correspond to states with equally spaced peaks. The dashed vertical line marks the BD point 
at $\rho=\rho_{\rm BD}\approx 1.80278$.

\begin{figure}[!t]
\centering
\includegraphics[scale=1]{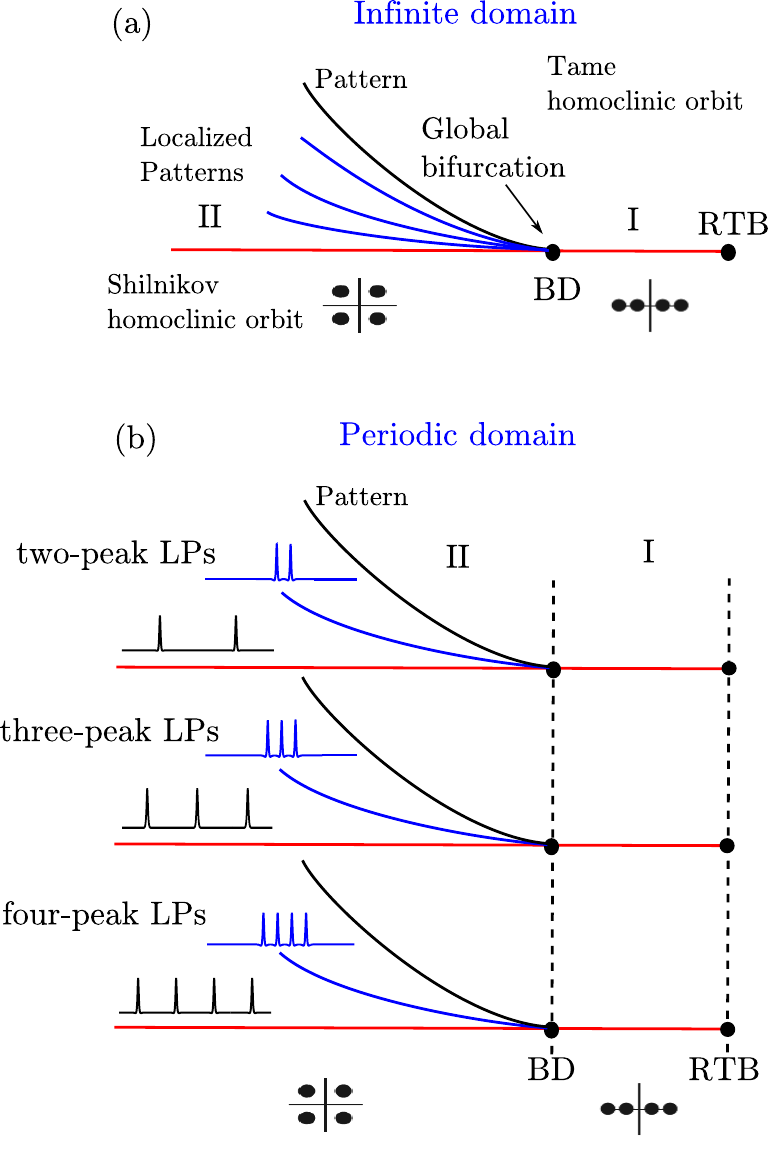}
\caption{(Color online) (a) Schematic unfolding of the global bifurcation occurring in the BD transition. At this bifurcation a single-peak LP with oscillatory tails changes into a single peak LS with monotonic tails (i.e., a spike) as $\rho$ increases. As this point is approached from the left the wavelength of the pattern involved in the formation of the LP states diverges. (b) Schematic unfolding of a single-peak LS in an infinite system (red line) into a large number of branches corresponding to states with different numbers of equispaced peaks. The LPs emerging from the global bifurcation at the BD point and connecting with the single-peak state in (a) are divided according to the number of peaks within each clump and are found in region II; region I contains equispaced structures organized in the foliated snaking structure generated in the RTB bifurcation at SN$_b$ as described in Section~\ref{sec:4b}. For $\theta=2.5$, $\rho_{\rm BD}\approx 1.80278$ while $\rho_b\approx 1.80501$ so region I is very 
narrow.}
\label{esquema_global}
\end{figure}

The LPs and the foliated snaking states are (nearly) degenerate in norm, which makes it difficult to discern the different type of bifurcation structures in Fig.~\ref{sepa}(a) from the norm alone since this only counts the number of peaks in a state regardless of their separation. However, the profiles of the different LPs shown in Fig.~\ref{sepa}(b)-(d) reveal that the LPs consist of clumped peaks whose separation $\Delta$, measured at half peak height, diverges at the BD point as $\rho$ increases. At this point, the branches of LPs disconnect and homoclinic snaking is destroyed. In fact in this regime the dominant evolution of the solution profile involves the separation of the peaks and not their amplitude. This makes the numerical computation of these branches a challenging task since a small change in $\rho$ results in a large change in $\Delta$ and the continuation fails unless $\rho$ is incremented by very small amounts. In contrast, for $\theta<2$ the separation between the peaks in an LP state is 
almost constant and of the order of the wavelength of the periodic pattern involved in their formation. The details of the wavelength selection mechanism within the snaking region in the Swift-Hohenberg equation are described in Ref.~\cite{Burke_Knobloch} and rely on the gradient structure the equation. No such theory exists for nongradient systems such as the LL equation. 

Below the BD transition LPs form by the heteroclinic tangle mechanism \cite{Beck}. The divergence of the separation $\Delta$ as the BD point is approached is the result of the divergence of the wavelength of the pattern involved in the tangle as $\rho\rightarrow\rho_{\rm BD}$. Thus, BD corresponds to a global bifurcation in space where the spatial period diverges, much as the signature of a global bifurcation in time is the divergence of an oscillation period \cite{Wiggins,Glendining}. This phenomenon is also known as the {\it blue sky catastrophe} \cite{Devaney_bluesky} or {\it wavelength blow-up} \cite{Fiedler}.

The unfolding of this global bifurcation on the real line is depicted in Fig.~\ref{esquema_global}(a). In region II LPs with multiple peaks
(i.e., multi-round homoclinic orbits) exist together with the periodic pattern (i.e., a limit cycle) involved in the heteroclinic tangle. The
wavelength of the pattern, and therefore the separation between the peaks, tends to infinity as the BD transition is approached and the 
bi-focus at $A_0^b$ becomes a saddle, resulting in the destruction of the Shilnikov-type homoclinic orbits present at lower 
$\rho$ \cite{Campneys_Edgar,Sandstede_hom,Belyakov_sinikov}.
In particular these LP structures bifurcate tangentially from the primary branch of homoclinic orbits and towards smaller $\rho$ (Region II).
In contrast, for $\rho>\rho_{\rm BD}$ the peaks have monotonic tails. Thus, by crossing this transition, a single-peak LP with oscillatory
tails (Shilnikov homoclinic orbit) existing in region II becomes a spike state with monotonic tails (tame homoclinic), the only type of LS 
that exist in region I \cite{Sandstede_hom,Campneys_Edgar}. The spike states may be thought of as arising from a one-parameter family of periodic orbits (i.e. patterns) as the spatial period diverges to infinity \cite{Devaney_A,Sandstede_hom,Belyakov_sinikov,Champneys_homoclinic}.

\begin{figure}[!t]
\centering
\includegraphics[scale=1]{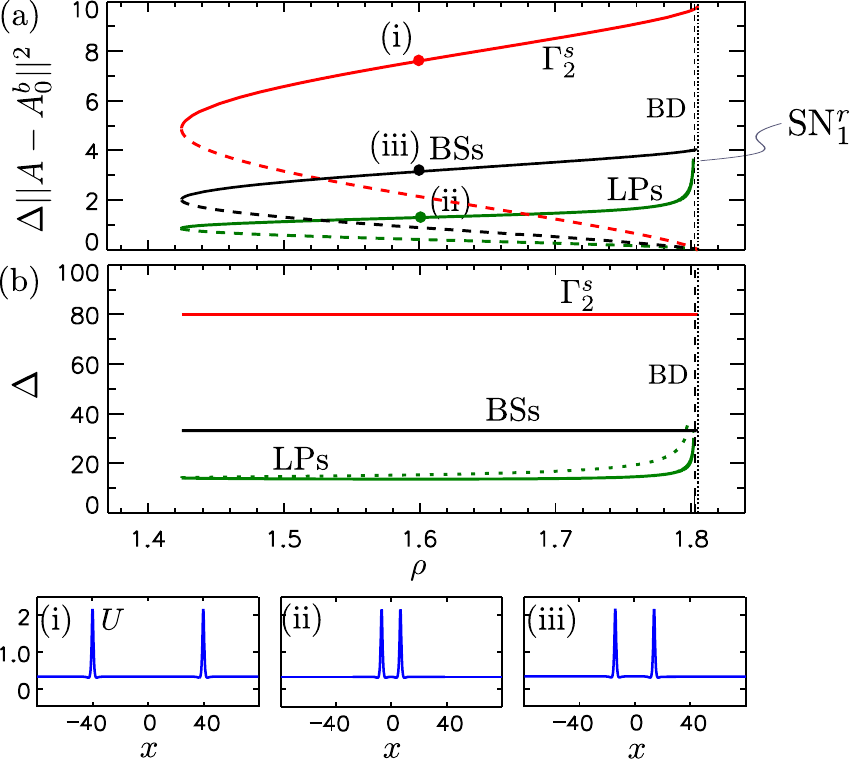}
\caption{(Color online) (a) Bifurcation diagram showing the reconnection of the two-peak LP branch at $\theta=2.5$ (green curve) arising from
homoclinic snaking at lower $\theta$ with the foliated snaking branches computed earlier in a global bifurcation at the BD transition (dashed line)
shown in terms of the quantity $\Delta\cdot||A-A^b_0||^2$ to separate the different solution branches. The branch $\Gamma_2$ of a periodic array of 
peaks, i.e., peaks separated by $L/2$, is shown in red. Bound states (BSs) with interpeak separation $\Delta$ between the minimum separation of 
order $2\pi/{\rm Im}[\lambda]$ and $L/2$ are shown in black; these form by the pairwise locking of oscillatory tails in the region $\rho<\rho_{\rm BD}$. 
Solid (dashed) lines indicate stable (unstable) states. (b) The separation $\Delta$ along the LS, $\Gamma_2$ and BS branches. The vertical dashed line correspond
to the BD transition, while the dotted one to the SN$_1^r$.
}
\label{sepa2}
\end{figure}

In our case, however, the system is periodic, and the global bifurcation at BD interacts with the foliated snaking skeleton computed on a finite periodic domain. Figure~\ref{sepa2} shows the solution branches corresponding to two-peak dissipative structures, both pattern and LPs, showing the product $\Delta\cdot||A-A^b_0||^2$ as a function of $\rho$. This new quantity allows us to separate the different branches that are otherwise not easily differentiated. The pattern state consists of two equispaced peaks while the LP state consists of a two-peak clump, periodically replicated on the real line. We see that, as $\rho$ increases, the branch corresponding to stable two-peak LPs (green) approaches $\Gamma_{2}^s$ and both branches eventually connect near the BD point when the separation of the peaks in the LP state reaches $\Delta=L/2$, the maximum separation that two peaks can reach in such a domain. The separation of the peaks increases with increasing $\rho$ along the lower, unstable LP branch as well, even 
as the peak amplitude tends to zero [dashed green line in Fig.~\ref{sepa}(d)], and also diverges as BD is approached. In both cases the divergence is a consequence of the vanishing of Im[$\lambda$] at BD.


In a similar way, LPs with three, four, or any given number of peaks undergo the same behavior and connect with branches $\Gamma_{3}$, $\Gamma_{4}$, etc., both stable and unstable, whose states correspond to the maximum separation that structures with three, four, etc., peaks can reach in a periodic system. Hence, the presence of foliated snaking unfolds the diagram sketched in Fig.~\ref{esquema_global}(a) resulting in the schematic picture in Fig.~\ref{esquema_global}(b), where we show the branches corresponding to structures with two, three and four peaks only.
\begin{figure*}[!t]
 \centering
  \includegraphics[scale=1]{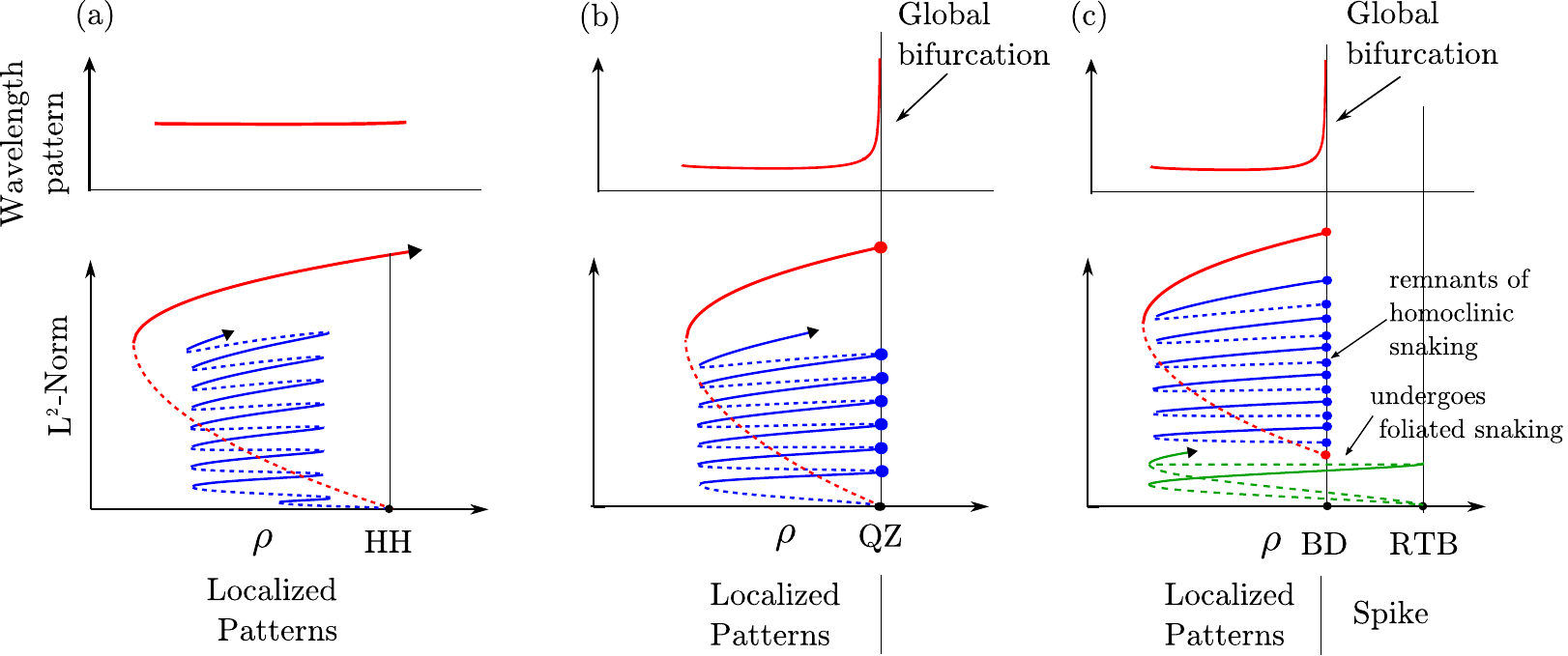}
  \caption{(Color online) Destruction of the homoclinic snaking. (a) $\theta<2$: LPs (blue) undergo homoclinic snaking, and the different families of solutions arise from the HH bifurcation. (b) $\theta=2$: The pattern (red) involved in the formation of the LPs undergoes a global bifurcation at the QZ point, where its wavelength becomes infinite. At this point the homoclinic snaking is destroyed. (c) $\theta>2$: A global bifurcation occurs at the BD point. At this point homoclinic snaking is destroyed. However, bright LS arise from the RTB bifurcation and undergo foliated snaking (green). The remnants of the homoclinic snaking branches (blue) reconnect with the branches forming the foliated snaking structure. The destruction of homoclinic snaking is indicated using {\color{blue}$\bullet$}.}
  \label{esquemaf}
\end{figure*}

Together with the previous structures, a large variety of bound states (BSs) of peaks can form via the locking of their oscillatory tails when $\rho<\rho_{\rm BD}$. By bound states we mean states consisting of two or more peaks separated by more than one wavelength of the coexisting periodic pattern. In such states the peaks interact very weakly but their separation is nonetheless locked to the wavelength in their oscillatory tails. In other words, these states consist of peaks separated by a distance greater than their minimum separation, i.e., the separation of the peaks in clumped LPs. A solution branch corresponding to one such state containing two pulses is shown in black in Fig.~\ref{sepa2}(a). Since the length $L$ of the domain is fixed the state cannot respond to small changes in the linear theory wavelength as $\rho$ and hence ${\rm Re}[\lambda]$ changes until the separation can no longer be maintained and the branch terminates in the vicinity of the BD point (Fig.~\ref{sepa2}(b)). In fact, careful 
computations show that the BS branch extends very slighly beyond BD (not shown), an effect we attribute to the effects of nonlinearity on the spatial separation of the two pulses, and we are led to conjecture that on the real line the two-pulse BS branch must in fact undergo a fold in $\rho>\rho_{\rm BD}$ at which it turns towards smaller $\rho$ before terminating at $\rho_{\rm BD}$. However, at the level of linear theory oscillatory tails are absent once $\rho>\rho_{\rm BD}$ and hence, modulo nonlinear effects, neither BSs nor LPs exist in Region I (Fig.~\ref{esquema_global}(b)). 
Both the LP and the BS states appear subcritically and so are initially unstable. However, both acquire stability at folds where they turn
towards larger $\rho$. Evidently at each $\rho_{\rm SN}<\rho<\rho_{\rm BD}$ there is a countably infinite number of two-peak BSs with different 
separations that accumulate on the primary Shilnikov homoclinic orbit as well as a countably infinite number of LP states with different numbers of peaks. Moreover, since the wavelength of the coexisting periodic state (equivalently $2\pi/{\rm Im}[\lambda]$) diverges as $\rho\to\rho_{\rm BD}$ from below so does the peak separation in every LP and in every BS.

\subsection{Transition between the previous scenarios}

The transition from region $41/30<\theta<2$ where homoclinic snaking is present to the scenario for $\theta>2$ where it is absent and reconnects with foliated snaking can be understood in terms of the periodic pattern involved in the heteroclinic tangle.

For $\theta<2$, see Fig.~\ref{esquemaf}(a), the LPs (blue) and the pattern (red) involved in their formation emerge together from the small amplitude bifurcation HH. At that point, the pattern has a characteristic wavelength of order $2\pi/k_c$, $k_c$ being the critical wavenumber, i.e., $k_c={\rm Im}[\lambda]$ at HH. Therefore, the LPs that form as a result of the heteroclinic tangle between this pattern and the bi-focus equilibrium $A_0^b$ will have a separation between peaks approximately equal to $2\pi/k_c$. Far from HH the LPs undergo homoclinic snaking, and multistability is found within the pinning or snaking region. As $\theta\rightarrow2$, $k_c\rightarrow0$, and the wavelength of the pattern involved in the formation of the LPs diverges. This first occurs at the QZ bifurcation that occurs at SN$_b$ when $\theta=2$ [Fig.~\ref{esquemaf}(b)] and the location of this divergence then moves down along $A_0^b$ as $\theta$ increases following the BD point. At QZ, the pattern becomes a tame homoclinic orbit, 
i.e., a 
single-peak LS with monotonic tails \cite{Devaney_A}. When $\theta$ increases past $\theta=2$ the BD transition separates from the RTB bifurcation at SN$_b$, and the tame homoclinic orbit and the pattern with infinite wavelength previously born together at QZ now arise from different bifurcations. This situation is depicted in Fig.~\ref{esquemaf}(c). Here small amplitude tame homoclinic orbits emerge from the RTB bifurcation (green), while a periodic pattern with an infinite wavelength bifurcates from the global homoclinic bifurcation occurring at the BD transition. To the left of BD LPs and the pattern involved in their formation coexist, and their bifurcation structure follows the remnants left over from homoclinic snaking [Fig.~\ref{esquemaf}(c)]. At the BD point there is a degenerate Shilnikov homoclinic orbit, while above BD the only structure that exists is a tame homoclinic orbit, and homoclinic snaking no longer exists. In periodic domains the single-peak LSs undergo foliated snaking and the branches 
remaining from homoclinic snaking reconnect with it.

\section{Conclusions}\label{sec:5}

In this work we have presented a detailed study of the bifurcation structure of bright dissipative structures arising in driven and dissipative optical
cavities in the anomalous GVD regime described by the LL equation in one spatial dimension. For $\theta<2$ two families of bright LPs arise from a HH
bifurcation (or equivalently an MI bifurcation), together with a spatially extended pattern with the critical wavenumber, and undergo homoclinic snaking
that continues until the available domain is filled and the LP states reconnect with the periodic pattern. These two families of solutions are interconnected
by unstable rung states consisting of drifting asymmetric LS. Together these branches constitute the snakes-and-ladders structure of the snaking or pinning 
region. For $\theta>2$ the HH bifurcation has turned into a BD transition from which small amplitude LPs no longer bifurcate. However, small amplitude LSs
still emerge from the RTB bifurcation at SN$_b$. Tracking this state through parameter space using 
numerical continuation allowed us to establish that it undergoes a bifurcation structure known as foliated snaking \cite{Ponedel_Knobloch}. This structure 
summarizes the global behavior of solutions consisting of $n$ equispaced peaks, $n=1,2,\dots$. All these branches originate together in the vicinity of 
SN$_b$, and with increasing amplitude each reconnects in a spatial period-halving bifurcation SR$_{2:1}$ with the branch of $2n$ identical peaks,
$n=1,2,\dots$. Moreover, the multi-peak states can be seen as periodic patterns in their own right, with wavelengths $L/2$, $L/4$, $L/8,...$ for the primary 
sequence, and $L/3$, $L/6$, $L/12...$ for the sequence starting with 3 peaks in the domain, and so on for $n=5,7,...$. Similar structures starting with
even integers $n\ne 2^p$ are also present. It follows that on the real line the presence of a fold in any of these branches is responsible for the 
presence spatially modulated multi-peak states, just as in the bistable Swift-Hohenberg equation where branches of 
``holes'' bifurcate from the vicinity of a fold of periodic states \cite{Bergeon}. We have not computed these states in this paper. On top of the
foliated snaking, solution branches corresponding to the LPs present for $\theta<2$ extend into the $\theta>2$ regime. These also consist of arrays
of peaks but these peaks are now clumped. The LP states are destroyed in a global homoclinic bifurcation at the BD transition 
\cite{Sandstede_hom,Campneys_Edgar}. One key feature of this bifurcation is the divergence of the wavelength of the pattern involved 
in the formation of the LPs as BD is approached, resulting in an abrupt growth in separation between the peaks constituting each LP state. 
As a result the stable and unstable branches corresponding to these states become disconnected, with each pair connecting to the corresponding equispaced state within the foliated snaking structure. This process is indicated in Fig \ref{esquema_global}(b): the LPs with 2 peaks connect at BD with the pair of states consisting of 2 peaks separated by L/2; the 3-peak LPs likewise connect with the equispaced 3-peak states, and so on (cf. Fig.~\ref{sepa}). A similar type of reconnection was in fact already observed in Ref.~\cite{Barashenkov} as well as in models of cellular and plant ecology \cite{Verschueren,Zelnik}. It is our understanding, however, that it has not yet been reported in any optical system such as that described in this paper.

The transition at $\theta=2$ between these two scenarios takes place via the QZ bifurcation which gives rise to all the relevant, local and global, bifurcations involved in the creation and destruction of LSs in this system. Despite its importance the unfolding of this point is still incompletely understood. This is because it corresponds to a codimension-two point with O(2) symmetry. Problems of this type are notoriously difficult although significant progress on other problems of this type have been made \cite{Dangelmayr,Rucklidge}. We have recently derived an amplitude equation valid in a neighborhood of this point similar to the Swift-Hohenberg equation with a quadratic nonlinearity appearing in Ref.~\cite{Buffoni}. A detailed study of this equation and its relevance in the present context will be presented elsewhere.

In the present paper we focused on steady states, although it is known that some of the structures we have described persist into regions where the LSs undergo oscillatory instabilities and even exhibit chaotic dynamics \cite{Leo_OE_2013,Parra-Rivas_PRA_KFCs,godey_stability_2014}. The consequences of these instabilities for the states we have described remain to be studied. The ultimate hope is that detailed studies of this kind will prove useful for the understanding of the formation and properties of frequency combs related to the underlying LSs as recently found in experiments using microcavities \cite{brasch_2016,okawachi_octave-spanning_2011,papp_microresonator_2014,ferdous_spectral_2011,herr_universal_2012,pfeifle_coherent_2014}.

\acknowledgments      
We acknowledge support from the Research Foundation--Flanders (FWO-Vlaanderen) (PPR), the Belgian Science Policy Office (BelSPO) under Grant IAP 7-35, the Research Council of the Vrije Universiteit Brussel, the Spanish MINECO and FEDER under grant ESOTECOS (FIS2015-63628-C2-1-R) (DG) and the National Science Foundation under grant DMS-1613132 (EK). We thank B. Ponedel for helpful discussions.

\section*{Appendix}
In this Appendix we calculate weakly nonlinear dissipative structures using multiple scale perturbation theory near the Hamiltonian-Hopf (HH) and the reversible Takens-Bogdanov (RTB) bifurcations in the Lugiato-Lefever (LL) equation keeping the detuning $\nu$ as a parameter. At the end of the calculation we set $\nu=1$ to obtain the corresponding results for the anomalous GVD regime of interest in this paper.

To study these types of solutions we write Eq.~(\ref{LLEsta}) in terms of the real and imaginary parts $U$ and $V$ of $A(x,t)$:
\begin{equation}
[\mathcal{L}+\mathcal{N}]\left[\begin{array}{c}
                                U \\ V
                               \end{array}\right]+\left[\begin{array}{c}
                                \rho \\ 0
                               \end{array}\right]=\left[\begin{array}{c}
                                0 \\ 0
                               \end{array}\right],
\end{equation}
with
\begin{equation}\label{lin_op_LL}
 \mathcal{L}=\left[\begin{array}{cc}
                    -1&\theta\\-\theta &-1
                   \end{array}
\right]+\left[\begin{array}{cc}
                    0&-\nu\\ \nu &0
                   \end{array}\right]\partial^2_x,
\end{equation}
and 
\begin{equation}
 \mathcal{N}=(U^2+V^2)\left[\begin{array}{cc}
                    0&-1\\1&0
                   \end{array}
\right]\left[\begin{array}{c}
               U\\V
              \end{array}
\right]\,.
\end{equation}

Following Ref.~\cite{BuYoKn} we fix the value of $\theta$ and suppose that in the neighborhood of such bifurcations the solutions are captured by the ansatz 
\begin{equation}\label{ansatz_HH_chap}
 \left[\begin{array}{c}
U\\V
\end{array}\right]= \left[\begin{array}{c}
U\\V
\end{array}\right]^*+ \left[\begin{array}{c}
u\\v
\end{array}\right],
\end{equation} 
where $U^*$ and $V^*$ represent the HSS $A_0^b$, and $u$ and $v$ capture the spatial dependence.


We next introduce appropriate asymptotic expansions for each variable in terms of a small parameter $\epsilon$ measuring the distance from the bifurcation point at $I_0=I_r$. Since $I_0$ depends on the parameter $\rho$ this distance translates into a departure of $\rho$ from the corresponding value $\rho_r$. The departure can be determined from a Taylor series expansion of $\rho(I_0)$ about $I_0=I_r$, where
\begin{equation}
\rho=\sqrt{I_0^3-2\theta I_0^2+(1+\theta^2)I_0}.
\end{equation}
Thus
\begin{multline}\label{taylor_rho}
 \rho(I_0)=\rho_r+\left(\frac{d\rho}{dI_0}\right)_{I_r}(I_0-I_r)+\\
 \frac{1}{2}\left(\frac{d^2\rho}{dI^2_0}\right)_{I_r}(I_0-I_r)^2+\mathcal{O}(3).
\end{multline}
For the RTB bifurcation the first derivative vanishes and one obtains
\begin{equation}
 \rho\approx\rho_b+\delta_b(I_0-I_b)^2.
\end{equation}
Writing $\epsilon=I_0-I_b$, we obtain
\begin{equation}
\epsilon=\sqrt{\frac{\rho-\rho_b}{\delta_b}},
\end{equation}
where
\begin{equation}\label{deltab1}
 \delta_b=\displaystyle\frac{1}{2}\left(\frac{d^2\rho}{dI^2_0}\right)_{I_b}.
\end{equation}
However, for the HH bifurcation the first derivative in the Taylor expansion (\ref{taylor_rho}) differs zero and thus
\begin{equation}
 \rho\approx\rho_c+\delta_c(I_0-I_c).
\end{equation}
In this case we write $\epsilon^2=I_0-I_c$, obtaining the expresion
\begin{equation}
\epsilon=\sqrt{\frac{\rho-\rho_c}{\delta_c}},
\end{equation}
where
\begin{equation}\label{deltac1}
 \delta_c=\left(\frac{d\rho}{dI_0}\right)_{I_c}.
\end{equation}
It follows that 
\begin{equation}
 \delta_b=-\frac{\sqrt{\theta^2-3}}{2\rho_b}<0,\label{deltab}
\end{equation}
and 
\begin{equation}
 \delta_c=\frac{(\theta-2)^2}{2\rho_c}>0.\label{deltac}
\end{equation}
Note that $\delta_b$ vanishes at the cusp bifurcation at $\theta=\sqrt{3}$ while $\delta_c$ vanishes at the QZ point $\theta=2$.

\subsection{Weakly nonlinear analysis around HH}\label{sec:pattern:2}

The HH point is defined for any value of $\theta<2$ by the condition $I_r=I_c=1$, or in terms of the energy injection $\rho$ by
\begin{equation}
 \rho_c=\sqrt{1+(\theta-1)^2}.
\end{equation}
In the following we fix the value of $\theta$ and, based on the scales defined above, approximate HSS by the series
\begin{equation}\label{eq.HSS}
 \left[\begin{array}{c}
U\\V
\end{array}\right]^*= \left[\begin{array}{c}
U_c\\V_c
\end{array}\right]+ \epsilon^2\left[\begin{array}{c}
U_2\\V_2
\end{array}\right]+...,
\end{equation}
and the $x$-dependent states by the series
\begin{equation}\label{space_depend}
 \left[\begin{array}{c}
u\\v
\end{array}\right]= \epsilon\left[\begin{array}{c}
u_1\\v_1
\end{array}\right]+ \epsilon^2\left[\begin{array}{c}
u_2\\v_2
\end{array}\right]+ \epsilon^3\left[\begin{array}{c}
u_3\\v_3
\end{array}\right]+...
\end{equation}

All quantities in (\ref{space_depend}) depend on both the short spatial scale $x$ and the long spatial scale $X\equiv\epsilon x$, i.e., $u_i=u_i(x,X(x))$ and $v_i=v_i(x,X(x))$. The differential operator $\partial^2_x$ acting on any of these fields thus takes the form
\begin{equation}
 \partial_x^2u_i(x,X(x))=\partial_x^2u_i+2\epsilon\partial_x\partial_Xu_i+\epsilon^2\partial_X^2u_i.
\end{equation}
It follows that the linear operator (\ref{lin_op_LL}) takes the form $\mathcal{L}=\mathcal{L}_0+\epsilon\mathcal{L}_1+\epsilon^2\mathcal{L}_2$,
where
\begin{subequations}
\begin{equation}
 \mathcal{L}_0=\left[\begin{array}{cc}
                    -1&\theta\\-\theta &-1
                   \end{array}
\right]+\left[\begin{array}{cc}
                    0&-\nu\\ \nu &0
                   \end{array}\right]\partial^2_x,
\end{equation}

\begin{equation}
 \mathcal{L}_1=2\left[\begin{array}{cc}
                    0&-\nu\\ \nu &0
                   \end{array}\right]\partial_{x}\partial_X,
\end{equation}

\begin{equation}
 \mathcal{L}_2=\left[\begin{array}{cc}
                    0&-\nu\\ \nu &0
                   \end{array}\right]\partial^2_X,
\end{equation} 
\end{subequations}
while the nonlinear terms take the form $\mathcal{N}=\mathcal{N}_0+\epsilon\mathcal{N}_1+\epsilon^2\mathcal{N}_2+\epsilon^3\mathcal{N}_3+\cdots$, where

\begin{equation}
 \mathcal{N}_0=(U_c^2+V_c^2)\left[\begin{array}{cc}
                    0&-1\\1&0
                   \end{array}
\right],
\end{equation}
\begin{equation}
 \mathcal{N}_1=2\left[\begin{array}{cc}
                    U_c & V_c
                   \end{array}
\right]\left[\begin{array}{c}
                    u_1\\v_1
                   \end{array}\right]\left[\begin{array}{cc}
                    0&-1\\1&0
                   \end{array}
\right],
\end{equation}
\begin{multline}
 \mathcal{N}_2=\left[\begin{array}{cc}
                               u_1 & v_1
                              \end{array}\right]\left[\begin{array}{c}
                               u_1\\v_1
                              \end{array}\right]
\left[\begin{array}{cc}
                    0&-1\\1&0
                   \end{array}
\right]+\\2\left[\begin{array}{cc}
                               U_c & V_c
                              \end{array}\right]\left[\begin{array}{c}
                               U_2+u_2\\V_2+v_2
                              \end{array}\right]
\left[\begin{array}{cc}
                    0&-1\\1&0
                   \end{array}
\right],
\end{multline}

\begin{multline}
\mathcal{N}_3=2\left[\begin{array}{cc}
                               U_c & V_c
                              \end{array}\right]\left[\begin{array}{c}
                               u_3\\v_3
                              \end{array}\right]
\left[\begin{array}{cc}
                    0&-1\\1&0
                   \end{array}
\right] 
                         \\+2\left[\begin{array}{cc}
                               u_1 & v_1
                              \end{array}\right]\left[\begin{array}{c}
                               U_2+u_2\\V_2+v_2
                              \end{array}\right]
\left[\begin{array}{cc}
                    0&-1\\1&0
                   \end{array}
\right].
\end{multline} 

At order $\epsilon^0$ stationary solutions satisfy
\begin{equation}
(\mathcal{L}_0+\mathcal{N}_0)\left[\begin{array}{c}
                              U_c\\V_c
                             \end{array}\right]+\left[\begin{array}{c}
                              \rho_c\\0
                             \end{array}\right]=\left[\begin{array}{c}
                              0\\0
                             \end{array}\right],
\end{equation}
and therefore
\begin{equation}
\left[\begin{array}{c}
U_c \\ V_c\end{array}\right]=\left[\begin{array}{c}
\displaystyle\frac{\rho_c}{1+(I_c-\theta)^2} \\ \displaystyle\frac{(I_c-\theta)\rho_c}{1+(I_c-\theta)^2}\end{array}\right]\,.
\end{equation}

At order $\epsilon^1$ we find that, 
\begin{equation}
(\mathcal{L}_0+\mathcal{N}_0)\left[\begin{array}{c}
                              u_1\\v_1
                             \end{array}\right] +\mathcal{N}_1\left[\begin{array}{c}
                              U_c\\V_c
                             \end{array}\right]=\left[\begin{array}{c}
                              0\\0
                             \end{array}\right],
\end{equation}
or 
\begin{equation}\label{eq.o1}
L\left[\begin{array}{c}
u_1 \\ v_1\end{array}\right]=\left[\begin{array}{c}
0 \\ 0\end{array}\right],
\end{equation}
where
\begin{multline}\label{LHH}
 L\equiv\mathcal{L}_0+\mathcal{N}_0+2\left[\begin{array}{cc}
                    0&-1\\1&0
                   \end{array}
\right]\left[\begin{array}{cc}
                    U_c^2&U_cV_c\\U_cV_c&V_c^2
                   \end{array}
\right]=\\ \left[\begin{array}{cc}
-(1+2U_cV_c) & \theta-I_c-2V_c^2-\nu\partial_x^2\\
-(\theta-I_c-2U_c^2-\nu\partial_x^2)& -1+2U_cV_c
\end{array}\right]
\end{multline}
is a singular differential operator.

To solve Eq.~\ref{LHH} we adopt the ansatz
\begin{equation}
\left[\begin{array}{c}
u_1 \\ v_1\end{array}\right]=\left[\begin{array}{c}a \\ b\end{array}\right]\left(\phi(X)e^{ik_cx}+\bar{\phi}(X)e^{-ik_cx}\right), 
\end{equation}
where $a,b\in\mathbb{R}$ and $\phi\in C^r(\mathbb{C})$. Nonzero solutions are present only when a certain solvability condition is satisfied. This condition demands that  
\begin{equation}\label{kc1_chapter}
k_c=\pm\sqrt{(2I_c-\theta)\nu\pm\sqrt{I_c^2-1}}
\end{equation}
and leads to the solution
\begin{equation}
\left[\begin{array}{c}
a \\ b\end{array}\right]=\left[\begin{array}{c}\displaystyle\frac{\theta-2V_c^2-I_c+ \nu k_c^2}{1+2U_cV_c}\\ 1\end{array}\right].\label{ab}
\end{equation}
Here $I_c$ is as yet undetermined but its value is found at next order in $\epsilon$. 

Proceeding to order $\epsilon^2$ we obtain
\begin{multline}
(\mathcal{L}_0+\mathcal{N}_0)\left[\begin{array}{c}
                              U_2+u_2\\V_2+v_2
                             \end{array}\right]+
                             (\mathcal{L}_1+\mathcal{N}_1)\left[\begin{array}{c}
                              u_1\\v_1
                             \end{array}\right]+\\
                             \mathcal{N}_2\left[\begin{array}{c}
                              U_c\\V_c
                             \end{array}\right]
                             +\left[\begin{array}{c}
                              \delta_c\\0
                             \end{array}\right]=\left[\begin{array}{c}
                              0\\0
                             \end{array}\right], 
\end{multline}
where $\delta_c$ is given by Eq.~(\ref{deltac}).

The space-independent terms can be written in the form
\begin{equation}
L\left[\begin{array}{c}
  U_2\\V_2
 \end{array}\right]+\left[\begin{array}{c}
  \delta_c\\0
 \end{array}\right]=\left[\begin{array}{c}
                              0\\0
                             \end{array}\right],
\end{equation}
whose solution takes the form 
\begin{equation}
 \left[\begin{array}{c}
U_2 \\ V_2\end{array}\right]=\delta_c d_0\left[\begin{array}{c}
1-2U_cV_c \\ 2U_c^2+I_c-\theta\end{array}\right]\equiv \delta_c \left[\begin{array}{c}
\tilde{U}_2\\ \tilde{V}_2\end{array}\right],
\end{equation}
where
\begin{equation}
 d_0=(1+\theta^2+3I_c^2-4\theta I_c)^{-1}.
\end{equation}
The remaining space-dependent terms are
\begin{multline}\label{second_order_space}
L\left[\begin{array}{c}
u_2 \\ v_2\end{array}\right]
+2\left[\begin{array}{cc}
0 & -\nu\\
\nu& 0
\end{array}\right]\partial_x\partial_X\left[\begin{array}{c}
u_1 \\ v_1\end{array}\right]+\\\\
\left[\begin{array}{cc}
-2u_1v_1& -(3v_1^2+u_1^2)\\
(3u_1^2+v_1^2) & 2v_1u_1
\end{array}\right]\left[\begin{array}{c}U_c \\ V_c\end{array}\right]=
\left[\begin{array}{c}
                              0\\0
                             \end{array}\right],
\end{multline}
which can be solved by using the ansatz
\begin{multline}\label{ansatzU2}
\left[\begin{array}{c}
u_2 \\ v_2\end{array}\right]=\left[\begin{array}{c}
A_0 \\
B_0
\end{array}\right]|\phi|^2
+ \left[\begin{array}{c}
A_1 \\ B_1\end{array}\right]\left(i\phi_Xe^{ik_cx}+ c.c.\right)+\\\left[\begin{array}{c}
A_2\\
B_2
\end{array}\right]\left(\phi^2e^{2ik_cx}+ c.c. \right).
\end{multline}
The coefficients $A_n$, $B_n$ follow from solutions of Eq.~(\ref{second_order_space}) of the form $e^{nik_cx}$, $n=0,1,2$, as described next.

When $n=0$ the $x$-derivatives in $L$ vanish. If we call the resulting operator $L_0$, the $n=0$ part of the solution is given by
\begin{equation}
\left[\begin{array}{c}
 A_0\\B_0
\end{array}\right]=2L_0^{-1}\left[\begin{array}{cc}
                            2a&3+a^2\\-(3a^2+1)&-2a
                           \end{array}\right]\left[\begin{array}{c}
                                       U_c\\V_c
                                      \end{array}\right]\,.
\end{equation}
Similarity, when $n=2$ we define an operator $L_2$ by replacing $\partial^2_x$ by $-4k_c^2$ and adding the result to $L_0$. Then
\begin{equation}
\left[\begin{array}{c}
 A_2\\B_2
\end{array}\right]=L_2^{-1}\left[\begin{array}{cc}
                            2a&3+a^2\\-(3a^2+1)&-2a
                           \end{array}\right]\left[\begin{array}{c}
                                       U_c\\V_c
                                      \end{array}\right]\,.
\end{equation}
When $n=1$, the operator $L_1$ is defined by replacing $\partial^2_x$ by $-k_c^2$ and adding the result to $L_0$. Then 
\begin{equation}\label{eq.n1}
 L_1\left[\begin{array}{c}
 A_1\\B_1
\end{array}\right]=-2k_c\left[\begin{array}{cc}
                          0 &- \nu\\ \nu & 0
                         \end{array}\right]\left[\begin{array}{c}
                                            a\\1
                                           \end{array}\right].
\end{equation}
However, this time $L_1$ is singular and the previous equation is only solvable if a solvability condition is satisfied. To find this condition we multiply this equation by the nullvector $\mathbf{w}$ of $L^{\dagger}$:
\begin{equation}
 \mathbf{w}=\left[\begin{array}{c}
w_1 \\
w_2
\end{array}\right]=\left[\begin{array}{c}
\displaystyle\frac{\theta-I_c-2U_c^2+\nu k_c^2}{-(1+2U_cV_c)} \\
1
\end{array}\right].
\end{equation}
The solvability condition is thus 
\begin{equation}
 w_1-a=0,
\end{equation}
which gives
\begin{equation}\label{kc2_chapter}
k_c^2=\nu(2I_c-\theta).
\end{equation}
Equation (\ref{kc1_chapter}) together with Eq.~(\ref{kc2_chapter}) yields the condition 
\begin{equation}\label{MI_chapter}
I_c=1
\end{equation}
for the HH bifurcation, as required. The two components of Eq.~(\ref{eq.n1}) are linearly related, and after multiplying both sides by $\left[\begin{array}{cc}
                 1 & 0 
                 \end{array}\right]$ we obtain
\begin{equation}
 B_1=-2k_c\displaystyle\frac{\left[\begin{array}{cc}
                 1 & 0 
                 \end{array}\right]\left[\begin{array}{cc}
                          0 &- \nu\\ \nu & 0
                         \end{array}\right]\left[\begin{array}{c}
                                            a\\1
                                           \end{array}\right]}{\left[\begin{array}{cc}
                 1 & 0 
                 \end{array}\right]L_1\left[\begin{array}{c}
                                     s \\ 1
                                    \end{array}\right]},
\end{equation}
where $s=A_1/B_1$ is arbitrary (subject to $s\neq a$). Without loss of generality, we choose $s=0$, i.e., $A_1=0$.

Finally, at order $\epsilon^3$ we obtain
\begin{multline}
(\mathcal{L}_0+\mathcal{N}_0)\left[\begin{array}{c}
                              u_3\\v_3
                             \end{array}\right]+
                             (\mathcal{L}_1+\mathcal{N}_1)\left[\begin{array}{c}
                              U_2+u_2\\V_2+v_2
                             \end{array}\right]+\\
                             (\mathcal{L}_2+\mathcal{N}_2)\left[\begin{array}{c}
                              u_1\\v_1
                             \end{array}\right]
                             +\mathcal{N}_3\left[\begin{array}{c}
                              U_c \\ V_c
                             \end{array}\right]=\left[\begin{array}{c}
                              0\\0
                             \end{array}\right].
\end{multline}
The solvability condition at this order is obtained by multiplying this equation by the nullvector of $L_1^{\dagger}$
\begin{equation}
 \mathbf{w}(x)=\zeta\left[\begin{array}{c}
                      w_1 \\ 1
                     \end{array}
\right]\left(e^{ik_cx}+e^{-ik_cx}\right),
\end{equation}
and integrating over $x$. The resulting equation can be written
\begin{equation}\label{eq.for.phi_HH}
 \nu C_1\phi_{XX}+\delta_c C_2\phi+C_3|\phi|^2\phi=0,
\end{equation}
where $\delta_c>0$ is given by (\ref{deltac}) and 
\begin{subequations}
\begin{equation}
 C_1=ak_cB_1,
\end{equation}
\begin{equation}
 C_2=aM_1+M_2,
\end{equation}
\begin{equation}
 C_3=aN_1+N_2,
\end{equation}
\end{subequations}
with
\begin{subequations}
 \begin{equation}
  M_1=-(V_c\tilde{U}_2+U_c\tilde{V}_2)a-(U_c\tilde{U}_2+3V_c\tilde{V}_2),
 \end{equation}
\begin{equation}
 M_2=(3U_c\tilde{U}_2+V_c\tilde{V}_2)a+V_c\tilde{U}_2+U_c\tilde{V}_2,
\end{equation}
\begin{equation}
 N_1=-(A_0+A_2)(U_c+aV_c)-(B_0+B_2)(aU_c+3V_c),
\end{equation}
\begin{equation}
 N_2=(B_0+B_2)(U_c+aV_c)+(A_0+A_2)(3aU_c+V_c).
\end{equation}
\end{subequations}
Evaluating the different coefficients we obtain
\begin{equation}
C_1=
-\frac{2 \, {\left(\theta^{2} - 2 \, \theta + 2\right)}}{\theta - 2}>0,
\end{equation}
\begin{equation}       	
C_2=
\frac{2 \, {\left(\theta^{2} - 2 \, \theta +
2\right)}^{\frac{3}{2}}}{{\left(\theta - 2\right)}^{4}}>0,
\end{equation}
\begin{equation}      	
C_3=\frac{4 \, {\left(\theta^{2} - 2 \, \theta +
2\right)}^{2} {\left(30 \, \theta - 41\right)}}{9 \, {\left(\theta -
2\right)}^{6}}.
\end{equation}

To solve Eq.~(\ref{eq.for.phi_HH}) we suppose that $\phi(X)=Be^{i\varphi}$, with $B\in\mathbb{R}^+$. With this ansatz two kinds of 
solutions can be found depending on whether $B$ depends on $X$ or not. If $B\neq B(X)$, then Eq.~(\ref{eq.for.phi_HH}) becomes 
\begin{equation}
\delta_c C_2B+C_3B^3=0, 
\end{equation}
with solutions $B_0=0$ and
$B_{\pm}=\pm\sqrt{-\delta_c C_2/ C_3}$, and the solution is 
\begin{equation}
\phi_{\rm P}=\sqrt{-\displaystyle\frac{\delta_c C_2}{C_3}}e^{i\varphi},
\end{equation}
where $\varphi$ in an arbitrary constant (due to translational invariance). This solution determines the spatially periodic state or pattern arising from HH.
The pattern can be sub- or supercritical depending on the value of $\theta$ with the transition between these two cases occurring at $\theta=41/30$ when $C_3=0$. If $C_3>0$, i.e., $\theta>41/30$, the pattern emerges subcritically, and supercritically otherwise.

In the subcritical regime, i.e., for $\theta>41/30$, localized patterns are also present and these can be found provided one supposes that $B=B(X)$. In this case Eq.~(\ref{eq.for.phi_HH}) becomes 
\begin{equation}
 \nu C_1 B_{XX}+\delta_c C_2B+C_3B^3=0,
\end{equation}
with the solution
\begin{equation}
 B(X)=\sqrt{\displaystyle\frac{-2\delta_c C_2}{C_3}}\,\textnormal{sech}\left(\displaystyle\sqrt{-\nu\delta_c C_2/C_1}(X-X_0)\right)
\end{equation}
centered at $X=X_0$. Hereafter we take $X_0=0$.
It follows that
\begin{equation}
 \phi_{\rm{LP}}(x)=\sqrt{\displaystyle\frac{-2\delta_c C_2}{C_3}}\,\textnormal{sech}\left(\displaystyle\sqrt{\frac{-\nu C_2(\rho-\rho_c)}{C_1}}x\right)e^{i\varphi}\,.
\end{equation}
Thus the leading order $x$-dependent solution arising at HH reads:
\begin{equation}\label{pattern_chapter}
\left[\begin{array}{c}
        u_1\\v_1
       \end{array}
\right]= 2\left[\begin{array}{c}
        a\\1
       \end{array}
\right]\displaystyle\phi\,\textnormal{cos}\left(k_cx\right)
\end{equation}
with $\phi=\phi_{\rm{P}}$ for the pattern P and $\phi=\phi_{\rm{LP}}$ for the localized pattern LP. Here $a$ is given by (\ref{ab}).

\subsection{Weakly nonlinear analysis around RTB}

The RTB point at the SN$_b$ is defined for any value of $\theta>2$ by the condition $I_r=I_b$ with
\begin{equation}
 I_b=\displaystyle\frac{1}{3}(2\theta-\sqrt{\theta^2-3}),
\end{equation}
or
\begin{equation}
 \rho_b=\sqrt{I_b^3-2\theta I_b^2+(1+\theta^2)I_b}.
\end{equation}

To find the solutions generated in this bifurcation we introduce appropriate asymptotic expansions for each variable as a function of $\epsilon$. For the HSS solutions we suppose that
\begin{equation}\label{eq.HSS_down}
 \left[\begin{array}{c}
U\\V
\end{array}\right]^*= \left[\begin{array}{c}
U_b\\V_b
\end{array}\right]+ \epsilon\left[\begin{array}{c}
U_1\\V_1
\end{array}\right]+ \epsilon^2\left[\begin{array}{c}
U_2\\V_2
\end{array}\right]+...,
\end{equation}
and for the space-dependent terms we take
\begin{equation}
 \left[\begin{array}{c}
u\\v
\end{array}\right]= \epsilon\left[\begin{array}{c}
u_1\\v_1
\end{array}\right]+ \epsilon^2\left[\begin{array}{c}
u_2\\v_2
\end{array}\right]+...,
\end{equation}
where $u_1$, $v_1$, $u_2$, and $v_2$ depend on the long scale variable $X\equiv\sqrt{\epsilon}x$. We also write $\rho=\rho_b+\delta_b\epsilon^2$.

In this case, the linear operator (\ref{lin_op_LL}) takes the form $\mathcal{L}=\mathcal{L}_0+\epsilon\mathcal{L}_1$
with 
\begin{subequations}
\begin{equation}
 \mathcal{L}_0=\left[\begin{array}{cc}
                    -1&\theta\\-\theta &-1
                   \end{array}
\right],
\end{equation}

\begin{equation}
 \mathcal{L}_1=\left[\begin{array}{cc}
                    0&-\nu\\ \nu &0
                   \end{array}\right]\partial^2_{X},
\end{equation}
\end{subequations}
and the nonlinear terms take the form $\mathcal{N}=\mathcal{N}_0+\epsilon\mathcal{N}_1+\epsilon^2\mathcal{N}_2+\cdots$, where
\begin{subequations}

\begin{equation}
 \mathcal{N}_0=(U_b^2+V_b^2)\left[\begin{array}{cc}
                    0&-1\\1&0
                   \end{array}
\right],
\end{equation}

\begin{equation}
 \mathcal{N}_1=2\left[\begin{array}{cc}
                       U_b&V_b
                      \end{array}
\right]\left[\begin{array}{c}
              U_1+u_1\\V_1+v_1
             \end{array}
\right]\left[\begin{array}{cc}
                    0&-1\\1&0
                   \end{array}
\right],
\end{equation}

\end{subequations}

\begin{multline}
 \mathcal{N}_2=\left[\begin{array}{cc}
                               U_1+u_1 & V_1+v_1
                              \end{array}\right]\left[\begin{array}{c}
                               U_1+u_1\\V_1+v_1
                              \end{array}\right]
\left[\begin{array}{cc}
                    0&-1\\1&0
                   \end{array}
\right]+\\
2\left[\begin{array}{cc}
                               U_b & V_b
                              \end{array}\right]\left[\begin{array}{c}
                               U_2+u_2\\V_2+v_2
                              \end{array}\right]
\left[\begin{array}{cc}
                    0&-1\\1&0
                   \end{array}
\right]\,.
\end{multline}

At order $\epsilon^0$ we therefore obtain
\begin{equation}
\left[\begin{array}{c}
U_b \\ V_b\end{array}\right]=\left[\begin{array}{c}
\displaystyle\frac{\rho_b}{1+(I_b-\theta)^2} \\ \displaystyle\frac{(I_b-\theta)\rho_b}{1+(I_b-\theta)^2}\end{array}\right].
\end{equation}
At order $\epsilon^1$
\begin{equation}
 (\mathcal{L}_0+\mathcal{N}_0)\left[\begin{array}{c}
                              U_1+u_1\\V_1+v_1
                             \end{array}\right] +\mathcal{N}_1\left[\begin{array}{c}
                              U_b\\V_b
                             \end{array}\right]=\left[\begin{array}{c}
                              0\\0
                             \end{array}\right]
\end{equation}
and this equation can be written as
\begin{equation}
 L\left[\begin{array}{c}
U_1+u_1\\V_1+v_1
\end{array}\right]=0,
\end{equation}
with the singular linear operator 
\begin{multline}
L\equiv\mathcal{L}_0+\mathcal{N}_0+2\left[\begin{array}{cc}
                    0&-1\\1&0
                   \end{array}
\right]\left[\begin{array}{cc}
                    U_b^2&U_bV_b\\U_bV_b&V_b^2
                   \end{array}
\right]=\\ =\left[\begin{array}{cc}
0 & 0\\
-(\theta-I_b-2U_b^2)& -2
\end{array}\right]\,.
\end{multline}

The $x$-independent equation has solutions that can be written in the form
\begin{equation}\label{s}
\left[\begin{array}{c}
U_1 \\ V_1\end{array}\right]=\mu\left[\begin{array}{c}1\\ \eta_b\end{array}\right],
\end{equation}
where
\begin{equation}
\eta_b=-\displaystyle\frac{1}{2}(\theta-I_b-2U_b^2)<0
\end{equation}
since $\theta>2$ and $\mu$ is obtained by solving the next order system. 

The general solution of the $x$-dependent equation gives
\begin{equation}
 \left[\begin{array}{c}
u_1 \\ v_1\end{array}\right]=\left[\begin{array}{c}
U_1 \\ V_1\end{array}\right]\phi(X),
\end{equation}
with $\phi(X)$ a function also determined at the next order.

Finally, at order $\epsilon^2$ one obtains 
\begin{multline}
 (\mathcal{L}_0+\mathcal{N}_0)\left[\begin{array}{c}
                              U_2+u_2\\V_2+v_2
                             \end{array}\right]+ (\mathcal{L}_1+\mathcal{N}_1)\left[\begin{array}{c}
                              U_1+u_1\\V_1+v_1
                             \end{array}\right]\\
                             +\mathcal{N}_2\left[\begin{array}{c}
                              U_b\\V_b
                             \end{array}\right]+\left[\begin{array}{c}
                              \delta_b\\0
                             \end{array}\right]=\left[\begin{array}{c}
                              0\\0
                             \end{array}\right]\,.
\end{multline}
The $x$-independent part of this equation gives
\begin{equation}
L\left[\begin{array}{c}
U_2 \\ V_2\end{array}\right] =
\left[\begin{array}{c}
2U_1V_1U_b +(2V_1^2+I_1)V_b-\delta_b\\
-(2U_1^2+I_1)U_b-2V_1U_1V_b
\end{array}\right],
\end{equation}\label{eq.2nd.orderHSS_down}
where $I_1\equiv U_1^2+V_1^2>0$, and $\delta_b<0$ is given by Eq.~(\ref{deltab}).
\begin{equation}\label{deltab2}
 \delta_b=-\frac{\sqrt{\theta^2-3}}{2\rho_b}<0.
\end{equation}

Since the operator $L$ is singular the above equation has no bounded solution unless a solvability condition is satisfied. This condition is given by 
\begin{equation}\label{mu_up}
\mu=\sqrt{-\delta_b}\mu_b,
\end{equation}
with
\begin{equation}\label{mu_up}
\mu_b\equiv-\sqrt{\frac{-1}{3\eta_b^2V_b+2\eta_b U_b+V_b}}<0.
\end{equation}

From the $x$-dependent terms we likewise find that
\begin{equation}\label{eq.2nd.spatial_down}
 L\left[\begin{array}{c}
u_2 \\ v_2\end{array}\right]=-\mathcal{P}_1\left[\begin{array}{c}
u_1 \\ v_1\end{array}\right]-\mathcal{P}_2\left[\begin{array}{c}
U_b \\ V_b\end{array}\right],
\end{equation}
with the linear operators
\begin{equation}
 \mathcal{P}_1=\left[\begin{array}{cc}
-(2U_bV_1+2U_1V_b) & -(\nu\partial_X^2+6V_bV_1+2U_bU_1)\\
\nu\partial_X^2+6U_bU_1+2V_bV_1& 2V_bU_1+2U_bV_1
\end{array}\right],
\end{equation}
and
\begin{equation}
 \mathcal{P}_2=\left[\begin{array}{cc}
-2v_1u_1 & -(3v_1^2+u_1^2)\\
3u_1^2+v_1^2 & 2v_1u_1
\end{array}\right].
\end{equation}
Because $L$ is singular, Eq.~(\ref{eq.2nd.spatial_down}) has no solution unless another solvability condition is satisfied. In the present case, this condition reads 
\begin{equation}\label{solvability.2_down}
 \left[\begin{array}{cc}
1 & 0\end{array}\right]\mathcal{P}_1\left[\begin{array}{c}
u_1 \\ v_1\end{array}\right]+ \left[\begin{array}{cc}
1 & 0\end{array}\right]\mathcal{P}_2\left[\begin{array}{c}
U_b \\ V_b\end{array}\right]=0\,.
\end{equation}

After some algebra, Eq.~(\ref{solvability.2_down}) reduces to an ordinary differential equation for $\phi(X)$,
\begin{equation}\label{psi.eq.1_down}
\nu c_1\phi_{XX}+c_2\phi+c_3\phi^2=0,
\end{equation}
where
\begin{equation}
\begin{array}{ccc}
c_1=-\displaystyle\frac{\mu_b\eta_b}{\sqrt{-\delta_b}}, & c_2=2, & c_3=1.
\end{array}
\end{equation}
Since $c_1<0$ for $\theta>2$ and $\delta_b<0$ this equation has the solution
\begin{equation}
 \phi(X)=-\frac{3}{2}\frac{c_2}{c_3}\textnormal{sech}^2\left(\displaystyle\frac{1}{2}\sqrt{-\frac{c_2\nu}{c_1}}(X-X_0)\right),
\end{equation}
equivalent to
\begin{equation}
\phi(X)=-3\,\textnormal{sech}^2\left(\displaystyle\frac{1}{2}\sqrt{\frac{2\nu\sqrt{-\delta_b}}{\mu_b\eta_b}}(X-X_0)\right),
\end{equation}
which represents a LS centered at $X=X_0$, hereafter at $X_0=0$. 
Since $X\equiv\sqrt{\epsilon}x$ and $\epsilon\equiv\sqrt{\displaystyle\frac{\rho-\rho_b}{\delta_b}}$ the corresponding spatial 
contribution to the first order solution in $\epsilon$ is given by
\begin{equation}
\phi(x)=-3\,{\textnormal{sech}}^2\left[ \displaystyle\frac{1}{2}\sqrt{\frac{2\nu}{\mu_b\eta_b}}(\rho_{b}-\rho)^{1/4}x\right],
\end{equation}
and the solution branch bifurcates towards smaller values of $\rho$.

\end{document}